\documentclass{psapm-l}

\newcommand{\ket}[1]{|#1\rangle}      % ket
\newcommand{\bra}[1]{\langle#1|}      % bra
\newcommand{\abs}[1]{|#1|}            % absolute value
\newcommand{\com}[2]{[#1, #2]}        % commutator
\newcommand{\acom}[2]{\{#1, #2\}}     % anticommutator
\newcommand{\set}[1]{\{#1\}}          % set
\newcommand{\logical}[1]{\overline{#1}} % logical operator/state
\newcommand{\order}[1]{|{#1}|}          % size of a set

\newcommand{\GF}{{\rm GF}(4)}          % the field GF(4)
\newcommand{\logzero}{\ket{\logical{0}}}  % logical 0 ket
\newcommand{\logone}{\ket{\logical{1}}}   % logical 1 ket
\newcommand{\logzerobra}{\bra{\logical{0}}} % logical 0 bra
\newcommand{\logonebra}{\bra{\logical{1}}}  % logical 1 bra
\newcommand{\E}{\mathcal{E}}            % the set of errors
\newcommand{\Hilbert}{\mathcal{H}}          % a Hilbert space
\newcommand{\Pauli}{\mathcal{P}}            % the Pauli group
\newcommand{\Clifford}{\mathcal{C}}         % the Clifford group

\newtheorem{thm}{Theorem}
\newtheorem{lemma}{Lemma}
\newtheorem{claim}{Claim}
\theoremstyle{definition}
\newtheorem{defn}{Definition}

\DeclareMathOperator{\tr}{tr}
\DeclareMathOperator{\wt}{wt}

\begin{document}

\title[Quantum Error Correction]{An Introduction to Quantum Error Correction and Fault-Tolerant Quantum Computation%
\footnote{Sections \ref{sec:intro}, \ref{sec:QECC}, and part of section~\ref{sec:moreQECC} of this chapter were originally published as~\cite{Gottesman:intro}.}}
\author{Daniel Gottesman}

\address{Perimeter Institute for Theoretical Physics \\
Waterloo, ON N2L 2Y5 \\
Canada}

\email{dgottesman@perimeterinstitute.ca}

\urladdr{http://www.perimeterinstitute.ca/personal/dgottesman/}


\begin{abstract}
Quantum states are very delicate, so it is likely some sort of
quantum error correction will be necessary to build reliable
quantum computers. The theory of quantum error-correcting codes
has some close ties to and some striking differences from the
theory of classical error-correcting codes. Many quantum codes can
be described in terms of the stabilizer of the codewords. The
stabilizer is a finite Abelian group, and allows a straightforward
characterization of the error-correcting properties of the code.
The stabilizer formalism for quantum codes also illustrates the
relationships to classical coding theory, particularly classical
codes over $\GF$, the finite field with four elements.  To build
a quantum computer which behaves correctly in the presence of errors,
we also need a theory of fault-tolerant quantum computation, instructing
us how to perform quantum gates on qubits which are encoded in a
quantum error-correcting code.  The threshold theorem states that
it is possible to create a quantum computer to perform an arbitrary
quantum computation provided the error rate per physical gate or
time step is below some constant threshold value.
\end{abstract}

\maketitle

\section{Background: the need for error correction}
\label{sec:intro}

Quantum computers have a great deal of potential, but to realize
that potential, they need some sort of protection from noise.

Classical computers don't use error correction.  One reason for
this is that classical computers use a large number of electrons,
so when one goes wrong, it is not too serious.  A single qubit in
a quantum computer will probably be just one, or a small number,
of particles, which already creates a need for some sort of error
correction.

Another reason is that classical computers are digital: after each
step, they correct themselves to the closer of 0 or 1.  Quantum
computers have a continuum of states, so it would seem, at first
glance, that they cannot do this. For instance, a likely source of
error is over-rotation: a state $\alpha \ket{0} + \beta \ket{1}$
might be supposed to become $\alpha \ket{0} + \beta e^{i\phi}
\ket{1}$, but instead becomes $\alpha \ket{0} + \beta e^{i(\phi +
\delta)} \ket{1}$.  The actual state is very close to the correct
state, but it is still wrong.  If we don't do something about
this, the small errors will build up over the course of the
computation, and eventually will become a big error.

Furthermore, quantum states are intrinsically delicate: looking at
one collapses it. $\alpha \ket{0} + \beta \ket{1}$ becomes
$\ket{0}$ with probability $\abs{\alpha}^2$ and $\ket{1}$ with
probability $\abs{\beta}^2$.  The environment is constantly trying
to look at the state, a process called {\em decoherence}.  One
goal of quantum error correction will be to prevent the
environment from looking at the data.

There is a well-developed theory of classical error-correcting
codes, but it doesn't apply here, at least not directly.  For one
thing, we need to keep the phase correct as well as correcting bit
flips.  There is another problem, too.  Consider the simplest
classical code, the repetition code:
\begin{align}
  0 & \rightarrow 000
\\1 & \rightarrow 111
\end{align}
It will correct a state such as $010$ to the majority value
(becoming $000$ in this case).\footnote{Actually, a classical
digital computer is using a repetition code -- each bit is encoded
in many electrons (the repetition), and after each time step, it
is returned to the value held by the majority of the electrons
(the error correction).}

We might try a quantum repetition code:
\begin{equation}
\ket{\psi} \rightarrow \ket{\psi} \otimes \ket{\psi} \otimes
\ket{\psi}
\end{equation}
However, no such code exists because of the No-Cloning
theorem~\cite{Dieks,WZ}:

\begin{thm}[No-Cloning]
There is no quantum operation that takes a state $\ket{\psi}$ to
$\ket{\psi} \otimes \ket{\psi}$ for all states $\ket{\psi}$.
\end{thm}

\begin{proof}
This fact is a simple consequence of the linearity of quantum
mechanics.  Suppose we had such an operation and $\ket{\psi}$ and
$\ket{\phi}$ are distinct. Then, by the definition of the
operation,
\begin{align}
\ket{\psi} & \rightarrow \ket{\psi} \ket{\psi}
\\ \ket{\phi} & \rightarrow \ket{\phi} \ket{\phi}
\\ \ket{\psi} + \ket{\phi} & \rightarrow
\left( \ket{\psi} + \ket{\phi} \right) \left( \ket{\psi} +
\ket{\phi} \right). \label{eq:cloned}
\end{align}
(Here, and frequently below, I omit normalization, which is
generally unimportant.)

But by linearity,
\begin{equation}
\ket{\psi} + \ket{\phi} \rightarrow \ket{\psi} \ket{\psi} +
\ket{\phi} \ket{\phi}.
\end{equation}
This differs from~(\ref{eq:cloned}) by the crossterm
\begin{equation}
\ket{\psi} \ket{\phi} + \ket{\phi} \ket{\psi}.
\end{equation}

\end{proof}

\section{Basic properties and structure of quantum error correction}
\label{sec:QECC}

\subsection{The nine-qubit code}

To solve these problems, we will try a variant of the repetition
code~\cite{Shor}.

\begin{align}
\ket{0} & \rightarrow \logzero = \left( \ket{000} + \ket{111}
\right) \left( \ket{000} + \ket{111} \right) \left( \ket{000} +
\ket{111} \right)
\\ \ket{1} & \rightarrow \logone = \left( \ket{000} -
\ket{111} \right) \left( \ket{000} - \ket{111} \right) \left(
\ket{000} - \ket{111} \right)
\end{align}

Note that this does not violate the No-Cloning theorem, since an
arbitrary codeword will be a linear superposition of these two
states
\begin{equation}
\alpha \logzero + \beta \logone \neq \left[ \alpha (\ket{000} +
\ket{111}) + \beta (\ket{000} - \ket{111}) \right]^{\otimes 3}.
\end{equation}
The superposition is linear in $\alpha$ and $\beta$.  The complete
set of codewords for this (or any other) quantum code form a
linear subspace of the Hilbert space, the {\em coding space}.

The inner layer of this code corrects bit flip errors:  We take
the majority within each set of three, so
\begin{equation}
\ket{010} \pm \ket{101} \rightarrow \ket{000} \pm \ket{111}.
\end{equation}
The outer layer corrects phase flip errors:  We take the majority
of the three signs, so
\begin{equation}
(\ket{\cdot} + \ket{\cdot}) (\ket{\cdot} - \ket{\cdot})
(\ket{\cdot} + \ket{\cdot}) \rightarrow (\ket{\cdot} +
\ket{\cdot}) (\ket{\cdot} + \ket{\cdot}) (\ket{\cdot} +
\ket{\cdot}).
\end{equation}
Since these two error correction steps are independent, the code
also works if there is both a bit flip error {\em and} a phase
flip error.

Note that in both cases, we must be careful to measure just what
we want to know and no more, or we would collapse the
superposition used in the code.  I'll discuss this in more detail
in section~\ref{sec:stabilizers}.

The bit flip, phase flip, and combined bit and phase flip errors
are important, so let's take a short digression to discuss them.
We'll also throw in the identity matrix, which is what we get if
no error occurs.  The definitions of these four operators are
given in table~\ref{tab:Pauli}.  The factor of $i$ in the
definition of $Y$ has little practical significance --- overall
phases in quantum mechanics are physically meaningless --- but it
makes some manipulations easier later.  It also makes some
manipulations harder, so either is a potentially reasonable
convention.

\begin{table}
\centering
\begin{tabular}{lll}
Identity & $I = \left( \begin{array}{cc} 1 & 0 \\ 0 & 1
\end{array} \right)$ & $I \ket{a} = \ket{a}$
\\ Bit Flip & $X = \left( \begin{array}{cc} 0 & 1 \\ 1 & 0
\end{array} \right)$ & $X \ket{a} = \ket{a \oplus 1}$
\\ Phase Flip & $Z = \left( \begin{array}{cc} 1 & 0 \\ 0 & -1
\end{array} \right)$ & $Z \ket{a} = (-1)^a \ket{a}$
\\ Bit \& Phase & $Y = \left( \begin{array}{cc} 0 & -i \\ i & 0
\end{array} \right) = iXZ$ & $Y \ket{a} = i (-1)^a \ket{a \oplus 1}$
\end{tabular}
\caption{The Pauli matrices} \label{tab:Pauli}
\end{table}

The group generated by tensor products of these $4$ operators is
called the Pauli group.  $X$, $Y$, and $Z$ anticommute: $XZ = -ZX$
(also written $\acom{X}{Z} = 0$).  Similarly, $\acom{X}{Y} = 0$
and $\acom{Y}{Z} = 0$.  Thus, the $n$-qubit Pauli group $\Pauli_n$
consists of the $4^n$ tensor products of $I$, $X$, $Y$, and $Z$,
and an overall phase of $\pm 1$ or $\pm i$, for a total of
$4^{n+1}$ elements.  The phase of the operators used is generally
not very important, but we can't discard it completely.  For
one thing, the fact that this is not an Abelian group is quite
important, and we would lose that if we dropped the phase!

$\Pauli_n$ is useful because of its nice algebraic properties. Any
pair of elements of $\Pauli_n$ either commute or anticommute.
Also, the square of any element of $\Pauli_n$ is $\pm 1$.  We
shall only need to work with the elements with square $+1$, which
are tensor products of $I$, $X$, $Y$, and $Z$ with an overall sign
$\pm 1$; the phase $i$ is only necessary to make $\Pauli_n$ a
group.  Define the {\em weight} $\wt(Q)$ of an operator $Q \in \Pauli_n$ to be
the number of tensor factors which are not $I$.  Thus, $X \otimes
Y \otimes I$ has weight $2$.

Another reason the Pauli matrices are important is that they span
the space of $2 \times 2$ matrices, and the $n$-qubit Pauli group
spans the space of $2^n \times 2^n$ matrices.  For instance, if we
have a general phase error
\begin{equation}
R_{\theta/2} = \left( \begin{array}{cc} 1 & 0 \\ 0 & e^{i\theta}
\end{array} \right) = e^{i \theta/2} \left( \begin{array}{cc}
e^{-i \theta/2} & 0 \\ 0 & e^{i \theta/2} \end{array} \right)
\end{equation}
(again, the overall phase does not matter), we can write it as
\begin{equation}
R_{\theta/2} = \cos \frac{\theta}{2} \ I - i \sin \frac{\theta}{2}
\ Z.
\end{equation}

It turns out that our earlier error correction procedure will also
correct this error, without any additional effort.  For instance,
the earlier procedure might use some extra qubits ({\em ancilla}
qubits) that are initialized to $\ket{0}$ and record what type of
error occurred. Then we look at the ancilla and invert the error
it tells us:
\begin{align}
Z \left( \alpha \logzero + \beta \logone \right) \otimes
\ket{0}_{\rm anc} & \rightarrow Z \left( \alpha \logzero + \beta
\logone \right) \otimes \ket{Z}_{\rm anc}
\\ & \rightarrow \left( \alpha \logzero + \beta
\logone \right) \otimes \ket{Z}_{\rm anc}
\\ I \left( \alpha \logzero + \beta \logone \right) \otimes
\ket{0}_{\rm anc} & \rightarrow I \left( \alpha \logzero + \beta
\logone \right) \otimes \ket{\text{no error}}_{\rm anc}
\\ & \rightarrow \left( \alpha \logzero + \beta
\logone \right) \otimes \ket{\text{no error}}_{\rm anc}
\end{align}

When the actual error is $R_{\theta/2}$, recording the error in
the ancilla gives us a superposition:
\begin{equation}
\cos \frac{\theta}{2} \ I \left( \alpha \logzero + \beta \logone
\right) \otimes \ket{\text{no error}}_{\rm anc} - i \sin
\frac{\theta}{2} \ Z \left( \alpha \logzero + \beta \logone
\right) \otimes \ket{Z}_{\rm anc}
\end{equation}
Then we measure the ancilla, which with probability $\sin^2
\theta/2$ gives us
\begin{equation}
Z \left( \alpha \logzero + \beta \logone \right) \otimes
\ket{Z}_{\rm anc},
\end{equation}
and with probability $\cos^2 \theta/2$ gives us
\begin{equation}
I \left( \alpha \logzero + \beta \logone \right) \otimes
\ket{\text{no error}}_{\rm anc}.
\end{equation}
In each case, inverting the error indicated in the ancilla
restores the original state.

It is easy to see this argument works for any linear combination
of errors~\cite{Shor,Steane:7}:

\begin{thm}
\label{thm:linear} If a quantum code corrects errors $A$ and $B$,
it also corrects any linear combination of $A$ and $B$.  In
particular, if it corrects all weight $t$ Pauli errors, then the
code corrects all $t$-qubit errors.
\end{thm}

So far, we have only considered individual unitary errors that
occur on the code.  But we can easily add in all possible quantum
errors.  The most general quantum operation, including
decoherence, interacts the quantum state with some extra qubits
via a unitary operation, then discards some qubits.  This process
can turn pure quantum states into mixed quantum states, which are
normally described using density matrices.  We can write the most
general operation as a transformation on density matrices
\begin{equation}
\rho \rightarrow \sum_i E_i \rho E_i^\dagger,
\end{equation}
where the $E_i$s are normalized so $\sum E_i^\dagger E_i = I$. The
density matrix $\rho$ can be considered to represent an ensemble
of pure quantum states $\ket{\psi}$, each of which, in this case,
should be in the coding space of the code.  Then this operation
simply performs the following operation on each $\ket{\psi}$:
\begin{equation}
\ket{\psi} \rightarrow E_i \ket{\psi} \text{ with probability }
\abs{E_i \ket{\psi}}^2.
\end{equation}
If we can correct each of the individual errors $E_i$, then we can
correct this general error as well.  For instance, for quantum
operations that only affect a single qubit of the code, $E_i$ will
necessarily be in the linear span of $I$, $X$, $Y$, and $Z$, so we
can correct it.  Thus, in the statement of
theorem~\ref{thm:linear}, ``all $t$-qubit errors'' really does
apply to {\em all} $t$-qubit errors, not just unitary ones.

We can go even further.  It is not unreasonable to expect that
every qubit in our nine-qubit code will be undergoing some small
error. For instance, qubit $i$ experiences the error $I + \epsilon
E_i$, where $E_i$ is some single-qubit error.  Then the overall
error is
\begin{equation}
\bigotimes (I + \epsilon E_i) = I + \epsilon \left(E_1 \otimes
I^{\otimes 8} + I \otimes E_2 \otimes I^{\otimes 7} + \ldots
\right) + O(\epsilon^2)
\end{equation}

That is, to order $\epsilon$, the actual error is the sum of
single-qubit errors, which we know the nine-qubit code can
correct.  That means that after the error correction procedure,
the state will be correct to $O(\epsilon^2)$ (when the two-qubit
error terms begin to become important).  While the code cannot
completely correct this error, it still produces a significant
improvement over not doing error correction when $\epsilon$ is
small.  A code correcting more errors would do even better.

\subsection{General properties of quantum error-correcting codes}

Let us try to understand what properties are essential to the
success of the nine-qubit code, and derive conditions for a
subspace to form a quantum error-correcting code (QECC).

One useful feature was {\em linearity}, which will be true of any
quantum code. We only need to correct a basis of errors ($I$, $X$,
$Y$, and $Z$ in the one-qubit case), and all other errors will
follow, as per theorem~\ref{thm:linear}.

In any code, we must never confuse $\logzero$ with $\logone$, even
in the presence of errors.  That is, $E \logzero$ is orthogonal to
$F \logone$:
\begin{equation}
\logzerobra E^\dagger F \logone = 0. \label{eq:orthog}
\end{equation}

It is {\em sufficient} to distinguish error $E$ from error $F$
when they act on $\logzero$ and $\logone$.  Then a measurement
will tell us exactly what the error is and we can correct it:
\begin{equation}
\logzerobra E^\dagger F \logzero = \logonebra E^\dagger F \logone
= 0 \label{eq:nondeg}
\end{equation}
for $E \neq F$.

But (\ref{eq:nondeg}) is not {\em necessary}: in the nine-qubit
code, we cannot distinguish between $Z_1$ and $Z_2$, but that is
OK, since we can correct either one with a single operation.  To
understand the necessary condition, it is helpful to look at the
operators $F_1 = (Z_1 + Z_2)/2$ and $F_2 = (Z_1 - Z_2)/2$ instead
of $Z_1$ and $Z_2$.  $F_1$ and $F_2$ span the same space as $Z_1$
and $Z_2$, so Shor's code certainly corrects them; let us try to
understand how. When we use the $F$s as the basis errors, now
equation~(\ref{eq:nondeg}) {\em is} satisfied.  That means we can
make a measurement and learn what the error is.  We also have to
invert it, and this is a potential problem, since $F_1$ and $F_2$
are not unitary.  However, $F_1$ acts the same way as $Z_1$ on the
coding space, so $Z_1^\dagger$ suffices to invert $F_1$ on the
states of interest.  $F_2$ acts the same way as the $0$ operator
on the coding space.  We can't invert this, but we don't need to
--- since $F_2$ annihilates codewords, it can never contribute a
component to the actual state of the system.

The requirement to invert the errors produces a third condition:
\begin{equation}
\logzerobra E^\dagger E \logzero = \logonebra E^\dagger E \logone.
\label{eq:equiv}
\end{equation}
Either this value is nonzero, as for $F_1$, in which case some
unitary operator will act the same way as $E$ on the coding space,
or it will be zero, as for $F_2$, in which case $E$ annihilates
codewords and never arises.

These arguments show that if there is some basis for the space of
errors for which equations (\ref{eq:orthog}), (\ref{eq:nondeg}),
and (\ref{eq:equiv}) hold, then the states $\logzero$ and
$\logone$ span a quantum error-correcting code.  Massaging these
three equations together and generalizing to multiple encoded
qubits, we get the following theorem~\cite{BDSW,KL}:

\begin{thm}
Suppose $\E$ is a linear space of errors acting on the Hilbert
space $\Hilbert$. Then a subspace $C$ of $\Hilbert$ forms a
quantum error-correcting code correcting the errors $\E$ iff
\begin{equation}
\bra{\psi} E^\dagger E \ket{\psi} = C(E) \label{eq:QECC}
\end{equation}
for all $E \in \E$.  The function $C(E)$ does not depend on the
state $\ket{\psi}$.
\end{thm}

\begin{proof}

Suppose $\set{E_a}$ is a basis for $\E$ and $\set{\ket{\psi_i}}$
is a basis for $C$.  By setting $E$ and $\ket{\psi}$ equal to the
basis elements and to the sum and difference of two basis elements
(with or without a phase factor $i$), we can see that
(\ref{eq:QECC}) is equivalent to
\begin{equation}
\bra{\psi_i} E_a^\dagger E_b \ket{\psi_j} = C_{ab} \delta_{ij},
\label{eq:QECC2}
\end{equation}
where $C_{ab}$ is a Hermitian matrix independent of $i$ and $j$.

Suppose equation~(\ref{eq:QECC2}) holds.  We can diagonalize
$C_{ab}$. This involves choosing a new basis $\set{F_a}$ for $\E$,
and the result is equations (\ref{eq:orthog}), (\ref{eq:nondeg}),
and (\ref{eq:equiv}).  The arguments before the theorem show that
we can measure the error, determine it uniquely (in the new
basis), and invert it (on the coding space). Thus, we have a
quantum error-correcting code.

Now suppose we have a quantum error-correcting code, and let
$\ket{\psi}$ and $\ket{\phi}$ be two distinct codewords.  Then we
must have
\begin{equation}
\bra{\psi} E^\dagger E \ket{\psi} = \bra{\phi} E^\dagger E
\ket{\phi}
\end{equation}
for all $E$.  That is, (\ref{eq:QECC}) must hold. If not, $E$
changes the relative size of $\ket{\psi}$ and $\ket{\phi}$.  Both
$\ket{\psi} + \ket{\phi}$ and $\ket{\psi} + c \ket{\phi}$ are
valid codewords, and
\begin{equation}
E (\ket{\psi} + \ket{\phi}) = N (\ket{\psi} + c \ket{\phi}),
\end{equation}
where $N$ is a normalization factor and
\begin{equation}
c = \bra{\psi} E^\dagger E \ket{\psi} / \bra{\phi} E^\dagger E
\ket{\phi}.
\end{equation}
The error $E$ will actually change the encoded state, which is a
failure of the code, unless $c=1$.

\end{proof}

There is a slight subtlety to the phrasing of
equation~(\ref{eq:QECC}).  We require $\E$ to be a linear space of
errors, which means that it must be closed under sums of errors
which may act on different qubits.  In contrast, for a code that
corrects $t$ errors, in~(\ref{eq:QECC2}), it is safe to consider
only $E_a$ and $E_b$ acting on just $t$ qubits.  We can restrict
even further, and only use Pauli operators as $E_a$ and $E_b$,
since they will span the space of $t$-qubit errors.  This leads us
to a third variation of the condition:
\begin{equation}
\bra{\psi} E \ket{\psi} = C'(E), \label{eq:QECC3}
\end{equation}
where $E$ is now any operator acting on $2t$ qubits (that is, it
replaces $E_a^\dagger E_b$ in (\ref{eq:QECC2})).  This can be
easily interpreted as saying that no measurement on $2t$ qubits
can learn information about the codeword.  Alternatively, it says
we can {\em detect} up to $2t$ errors on the code without
necessarily being able to say what those errors are.  That is, we
can distinguish those errors from the identity.

If the matrix $C_{ab}$ in (\ref{eq:QECC2}) has maximum rank, the
code is called {\em nondegenerate}.  If not, as for the nine-qubit
code, the code is {\em degenerate}.  In a degenerate code,
different errors look the same when acting on the coding subspace.

For a nondegenerate code, we can set a simple bound on the
parameters of the code simply by counting states.  Each error $E$
acting on each basis codeword $\ket{\psi_i}$ produces a linearly
independent state.  All of these states must fit in the full
Hilbert space of $n$ qubits, which has dimension $2^n$.  If the
code encodes $k$ qubits, and corrects errors on up to $t$ qubits,
then
\begin{equation}
\left( \sum_{j=0}^t 3^j \binom{n}{j} \right) 2^k \leq 2^n.
\label{eq:QHB}
\end{equation}
The quantity in parentheses is the number of errors of {\em
weight} $t$ or less: that is, the number of tensor products of
$I$, $X$, $Y$, and $Z$ that are the identity in all but $t$ or
fewer places.  This inequality is called the {\em quantum Hamming
bound}.  While the quantum Hamming bound only applies to
nondegenerate codes, we do not know of any codes that beat it.

For $t=1$, $k=1$, the quantum Hamming bound tells us $n \geq 5$.
In fact, there is a code with $n=5$, which you will see later.  A
code that corrects $t$ errors is said to have {\em distance}
$2t+1$, because it takes $2t+1$ single-qubit changes to get from
one codeword to another.  We can also define distance as the
minimum weight of an operator $E$ that violates equation
(\ref{eq:QECC3}) (a definition which also allows codes of even
distance). A quantum code using $n$ qubits to encode $k$ qubits
with distance $d$ is written as an $[[n, k, d]]$ code (the double
brackets distinguish it from a classical code). Thus, the
nine-qubit code is a $[[9, 1, 3]]$ code, and the five-qubit code
is a $[[5, 1, 3]]$ code.

We can also set a lower bound telling us when codes exist.  I will
not prove this here, but an $[[n, k, d]]$ code exists when
\begin{equation}
\left( \sum_{j=0}^{d-1} 3^j \binom{n}{j} \right) 2^k \leq 2^n
\label{eq:QGV}
\end{equation}
(known as the quantum Gilbert-Varshamov bound~\cite{CRSS}).  This
differs from the quantum Hamming bound in that the sum goes up to
$d-1$ (which is equal to $2t$) rather than stopping at $t$.

\begin{thm}
A quantum $[[n, k, d]]$ code exists when (\ref{eq:QGV})~holds. Any
nondegenerate $[[n, k, d]]$ code must satisfy~(\ref{eq:QHB}).  For
large $n$, $R = k/n$ and $p = d/2n$ fixed, the best nondegenerate
quantum codes satisfy
\begin{equation}
1 - 2p \log_2 3 - H(2p) \leq R \leq 1 - p \log_2 3 - H(p),
\end{equation}
where $H(x) = - x \log_2 x - (1-x) \log_2 (1-x)$.
\end{thm}

One further bound, known as the Knill-Laflamme bound~\cite{KL} or
the quantum Singleton bound, applies even to degenerate quantum
codes. For an $[[n,k,d]]$ quantum code,
\begin{equation}
n - k \geq 2d - 2.
\end{equation}
This shows that the $[[5,1,3]]$ code really is optimal --- a
$[[4,1,3]]$ code would violate this bound.

I will not prove the general case of this bound, but the case of
$k=1$ can be easily understood as a consequence of the No-Cloning
theorem.  Suppose $r$ qubits of the code are missing.  We can
substitute $\ket{0}$ states for the missing qubits, but there are
$r$ errors on the resulting codeword.  The errors are of unknown
type, but all the possibilities are on the same set of $r$ qubits.
Thus, all products $E_a^\dagger E_b$ in condition~(\ref{eq:QECC2})
have weight $r$ or less, so this sort of error (an ``erasure''
error~\cite{GBP}) can be corrected by a code of distance $r+1$.
Now suppose we had an $[[n, 1, d]]$ code with $n \leq 2d-2$.  Then
we could split the qubits in the code into two groups of size at
most $d-1$. Each group would have been subject to at most $d-1$
erasure errors, and could therefore be corrected without access to
the other group. This would produce two copies of the encoded
state, which we know is impossible.

\subsection{Stabilizer codes}
\label{sec:stabilizers}

Now let us return to the nine-qubit code, and examine precisely
what we need to do to correct errors.

First, we must determine if the first three qubits are all the
same, and if not, which is different.  We can do this by measuring
the parity of the first two qubits and the parity of the second
and third qubits.  That is, we measure
\begin{equation}
Z \otimes Z \otimes I \text{ and } I \otimes Z \otimes Z.
\end{equation}
The first tells us if an $X$ error has occurred on qubits one or
two, and the second tells us if an $X$ error has occurred on
qubits two or three.  Note that the error detected in both cases
anticommutes with the error measured.  Combining the two pieces of
information tells us precisely where the error is.

We do the same thing for the other two sets of three.  That gives
us four more operators to measure.  Note that measuring $Z \otimes
Z$ gives us just the information we want and no more.  This is
crucial so that we do not collapse the superpositions used in the
code.  We can do this by bringing in an ancilla qubit.  We start
it in the state $\ket{0} + \ket{1}$ and perform controlled-$Z$
operations to the first and second qubits of the code:
\begin{align}
\left( \ket{0} + \ket{1} \right) \sum_{abc} c_{abc} \ket{abc} &
\rightarrow \sum_{abc} c_{abc} \left( \ket{0} \ket{abc} + (-1)^{a
\oplus b} \ket{1} \ket{abc} \right)
\\ & = \sum_{abc} c_{abc} \left( \ket{0} + (-1)^{{\rm parity} (a,b)}
\ket{1} \right) \ket{abc}.
\end{align}
At this point, measuring the ancilla in the basis $\ket{0} \pm
\ket{1}$ will tell us the eigenvalue of $Z \otimes Z \otimes I$,
but nothing else about the data.

Second, we must check if the three signs are the same or
different.  We do this by measuring
\begin{equation}
X \otimes X \otimes X \otimes X \otimes X \otimes X \otimes I
\otimes I \otimes I
\end{equation}
and
\begin{equation}
I \otimes I \otimes I \otimes X \otimes X \otimes X \otimes X
\otimes X \otimes X.
\end{equation}
This gives us a total of $8$ operators to measure.  These two
measurements detect $Z$ errors on the first six and last six
qubits, correspondingly.  Again note that the error detected
anticommutes with the operator measured.

This is no coincidence: in each case, we are measuring an operator
$M$ which should have eigenvalue $+1$ for any codeword:
\begin{equation}
M \ket{\psi} = \ket{\psi}.
\end{equation}
If an error $E$ which anticommutes with $M$ has occurred, then the
true state is $E \ket{\psi}$, and
\begin{equation}
M \left( E \ket{\psi} \right) = - E M \ket{\psi} = - E \ket{\psi}.
\end{equation}
That is, the new state has eigenvalue $-1$ instead of $+1$.  We
use this fact to correct errors: each single-qubit error $E$
anticommutes with a particular set of operators $\set{M}$; which
set, exactly, tells us what $E$ is.

In the case of the nine-qubit code, we cannot tell exactly what
$E$ is, but it does not matter.  For instance, we cannot
distinguish $Z_1$ and $Z_2$ because
\begin{equation}
Z_1 Z_2 \ket{\psi} = \ket{\psi} \ \ \Longleftrightarrow \ \ Z_1
\ket{\psi} = Z_2 \ket{\psi}.
\end{equation}
This is an example of the fact that the nine-qubit code is
degenerate.

Table~\ref{tab:nine} summarizes the operators we measured.
\begin{table}
\centering
\begin{tabular}{ccccccccc}
   $Z$ & $Z$ & $I$ & $I$ & $I$ & $I$ & $I$ & $I$ & $I$
\\ $I$ & $Z$ & $Z$ & $I$ & $I$ & $I$ & $I$ & $I$ & $I$
\\ $I$ & $I$ & $I$ & $Z$ & $Z$ & $I$ & $I$ & $I$ & $I$
\\ $I$ & $I$ & $I$ & $I$ & $Z$ & $Z$ & $I$ & $I$ & $I$
\\ $I$ & $I$ & $I$ & $I$ & $I$ & $I$ & $Z$ & $Z$ & $I$
\\ $I$ & $I$ & $I$ & $I$ & $I$ & $I$ & $I$ & $Z$ & $Z$
\\ $X$ & $X$ & $X$ & $X$ & $X$ & $X$ & $I$ & $I$ & $I$
\\ $I$ & $I$ & $I$ & $X$ & $X$ & $X$ & $X$ & $X$ & $X$
\end{tabular}
\caption{The stabilizer for the nine-qubit code.  Each column
represents a different qubit.} \label{tab:nine}
\end{table}
These $8$ operators generate an Abelian group called the {\em
stabilizer} of the nine-qubit code.  The stabilizer contains all
operators $M$ in the Pauli group for which $M \ket{\psi} =
\ket{\psi}$ for all $\ket{\psi}$ in the code.

Conversely, given an Abelian subgroup $S$ of the Pauli group
$\Pauli_n$ (which, if you recall, consists of tensor products of
$I$, $X$, $Y$, and $Z$ with an overall phase of $\pm 1, \pm i$),
we can define a quantum code $T(S)$ as the set of states
$\ket{\psi}$ for which $M \ket{\psi} = \ket{\psi}$ for all $M \in
S$.  $S$ must be Abelian and cannot contain $-1$, or the code is
trivial: If $M, N \in S$,
\begin{align}
MN \ket{\psi} = M \ket{\psi} & = \ket{\psi}
\\ NM \ket{\psi} = N \ket{\psi} & = \ket{\psi}
\end{align}
so
\begin{equation}
\com{M}{N} \ket{\psi} = MN \ket{\psi} - NM \ket{\psi} = 0.
\end{equation}
Since elements of the Pauli group either commute or anticommute,
$\com{M}{N} = 0$.  Clearly, if $M = -1 \in S$, there is no
nontrivial $\ket{\psi}$ for which $M \ket{\psi} = \ket{\psi}$.

If these conditions are satisfied, there will be a nontrivial
subspace consisting of states fixed by all elements of the
stabilizer.  We can tell how many errors the code corrects by
looking at operators that commute with the stabilizer. We can
correct errors $E$ and $F$ if either $E^\dagger F \in S$ (so $E$
and $F$ act the same on codewords), or if $\exists M \in S\ {\rm
s.t.}\ \acom{M}{E^\dagger F} = 0$, in which case measuring the
operator $M$ distinguishes between $E$ and $F$. If the first
condition is ever true, the stabilizer code is degenerate;
otherwise it is nondegenerate.

We can codify this by looking at the normalizer $N(S)$ of $S$ in
the Pauli group (which is in this case equal to the centralizer,
composed of Pauli operators which commute with $S$).  The distance
$d$ of the code is the minimum weight of any operator in $N(S)
\setminus S$~\cite{CRSS,Gottesman}.

\begin{thm}
\label{thm:stabilizer}
Let $S$ be an Abelian subgroup of order $2^a$ of the $n$-qubit
Pauli group, and suppose $-1 \not\in S$.  Let $d$ be the minimum
weight of an operator in $N(S) \setminus S$.  Then the space of
states $T(S)$ stabilized by all elements of $S$ is an $[[n, n-a,
d]]$ quantum code.
\end{thm}

To correct errors of weight $(d-1)/2$ or below, we simply measure
the generators of $S$.  This will give us a list of eigenvalues,
the {\em error syndrome}, which tells us whether the error $E$
commutes or anticommutes with each of the generators.  The error
syndromes of $E$ and $F$ are equal iff the error syndrome of
$E^\dagger F$ is trivial.  For a nondegenerate code, the error
syndrome uniquely determines the error $E$ (up to a trivial
overall phase) --- the generator that anticommutes with $E^\dagger
F$ distinguishes $E$ from $F$. For a degenerate code, the error
syndrome is not unique, but error syndromes are only repeated when
$E^\dagger F \in S$, implying $E$ and $F$ act the same way on the
codewords.

If the stabilizer has $a$ generators, then the code encodes $n-a$
qubits.  Each generator divides the allowed Hilbert space into
$+1$ and $-1$ eigenspaces of equal sizes. To prove the statement,
note that we can find an element $G$ of the Pauli group that has
any given error syndrome (though $G$ may have weight greater than
$(d-1)/2$, or even greater than $d$). Each $G$ maps $T(S)$ into an
orthogonal but isomorphic subspace, and there are $2^a$ possible
error syndromes, so $T(S)$ has dimension at most $2^n/2^a$.  In
addition, the Pauli group spans $U(2^n)$, so its orbit acting on
any single state contains a basis for $\Hilbert$. Every Pauli
operator has {\em some} error syndrome, so $T(S)$ has dimension
exactly $2^{n-a}$.

\section{More quantum error-correcting codes and their structure}
\label{sec:moreQECC}

\subsection{Some other important codes}
\label{sec:codes}

Stabilizers make it easy to describe new codes.  For instance, we
can start from classical coding theory, which describes a linear
code by a generator matrix or its dual, the parity check matrix.
Each row of the generator matrix is a codeword, and the other
codewords are all linear combinations of the rows of the generator
matrix.  The rows of the parity check matrix specify parity checks
all the classical codewords must satisfy.  (In quantum codes, the
stabilizer is closely analogous to the classical parity check
matrix.)  One well-known code is the seven-bit Hamming code
correcting one error, with parity check matrix
\begin{equation}
\left( \begin{array}{ccccccc}
   1 & 1 & 1 & 1 & 0 & 0 & 0
\\ 1 & 1 & 0 & 0 & 1 & 1 & 0
\\ 1 & 0 & 1 & 0 & 1 & 0 & 1
\end{array} \right).
\end{equation}

If we replace each $1$ in this matrix by the operator $Z$, and $0$
by $I$, we are really changing nothing, just specifying three
operators that implement the parity check measurements.  The
statement that the classical Hamming code corrects one error is
the statement that each bit flip error of weight one or two
anticommutes with one of these three operators.

Now suppose we replace each $1$ by $X$ instead of $Z$.  We again
get three operators, and they will anticommute with any weight one
or two $Z$ error.  Thus, if we make a stabilizer out of the three
$Z$ operators and the three $X$ operators, as in
table~\ref{tab:seven}, we get a code that can correct any single
qubit error~\cite{Steane:7}.  $X$ errors are picked up by the
first three generators, $Z$ errors by the last three, and $Y$
errors are distinguished by showing up in both halves.  Of course,
there is one thing to check: the stabilizer must be Abelian; but
that is easily verified.
\begin{table}
\centering
\begin{tabular}{ccccccc}
   $Z$ & $Z$ & $Z$ & $Z$ & $I$ & $I$ & $I$
\\ $Z$ & $Z$ & $I$ & $I$ & $Z$ & $Z$ & $I$
\\ $Z$ & $I$ & $Z$ & $I$ & $Z$ & $I$ & $Z$
\\ $X$ & $X$ & $X$ & $X$ & $I$ & $I$ & $I$
\\ $X$ & $X$ & $I$ & $I$ & $X$ & $X$ & $I$
\\ $X$ & $I$ & $X$ & $I$ & $X$ & $I$ & $X$
\end{tabular}
\caption{Stabilizer for the seven-qubit code.} \label{tab:seven}
\end{table}
The stabilizer has $6$ generators on $7$ qubits, so it encodes $1$
qubit --- it is a $[[7, 1, 3]]$ code.

In this example, we used the same classical code for both the $X$
and $Z$ generators, but there was no reason we had to do so.  We
could have used any two classical codes $C_1$ and
$C_2$~\cite{CS,Steane}. The only requirement is that the $X$ and
$Z$ generators commute. This corresponds to the statement that
$C_2^\perp \subseteq C_1$ ($C_2^\perp$ is the dual code to $C_2$,
consisting of those words which are orthogonal to the codewords of
$C_2$).  If $C_1$ is an $[n, k_1, d_1]$ code, and $C_2$ is an $[n,
k_2, d_2]$ code (recall single brackets means a classical code),
then the corresponding quantum code is an $[[n, k_1 + k_2 - n,
\min(d_1, d_2)]]$ code.\footnote{In fact, the true distance of the
code could be larger than expected because of the possibility of
degeneracy, which would not have been a factor for the classical
codes.} This construction is known as the CSS construction after
its inventors Calderbank, Shor, and Steane.

The codewords of a CSS code have a particularly nice form.  They
all must satisfy the same parity checks as the classical code
$C_1$, so all codewords will be superpositions of words of $C_1$.
The parity check matrix of $C_2$ is the generator matrix of
$C_2^\perp$, so the $X$ generators of the stabilizer add a word of
$C_2^\perp$ to the state.  Thus, the codewords of a CSS code are
of the form
\begin{equation}
\sum_{w \in C_2^\perp} \ket{u + w},
\end{equation}
where $u \in C_1$ ($C_2^\perp \subseteq C_1$, so $u + w \in C_1$).
If we perform a Hadamard transform
\begin{align}
   \ket{0} & \longleftrightarrow \ket{0} + \ket{1}
\\ \ket{1} & \longleftrightarrow \ket{0} - \ket{1}
\end{align}
on each qubit of the code, we switch the $Z$ basis with the $X$
basis, and $C_1$ with $C_2$, so the codewords are now
\begin{equation}
\sum_{w \in C_1^\perp} \ket{u + w} \quad (u \in C_2).
\end{equation}
Thus, to correct errors for a CSS code, we can measure the
parities of $C_1$ in the $Z$ basis, and the parities of $C_2$ in
the $X$ basis.

Another even smaller quantum code is the $[[5, 1, 3]]$ code I
promised earlier~\cite{BDSW,LMPZ}.  Its stabilizer is given in
table~\ref{tab:five}.
\begin{table}
\centering
\begin{tabular}{ccccc}
   $X$ & $Z$ & $Z$ & $X$ & $I$
\\ $I$ & $X$ & $Z$ & $Z$ & $X$
\\ $X$ & $I$ & $X$ & $Z$ & $Z$
\\ $Z$ & $X$ & $I$ & $X$ & $Z$
\end{tabular}
\caption{The stabilizer for the five-qubit code.} \label{tab:five}
\end{table}
I leave it to you to verify that it commutes and actually does
have distance $3$.  You can also work out the codewords.  Since
multiplication by $M \in S$ merely rearranges elements of the
group $S$, the sum
\begin{equation}
\left( \sum_{M \in S} M \right) \ket{\phi} \label{eq:proj}
\end{equation}
is in the code for any state $\ket{\phi}$.  You only need find two
states $\ket{\phi}$ for which (\ref{eq:proj})~is nonzero.  Note
that as well as telling us about the error-correcting properties
of the code, the stabilizer provides a more compact notation for
the coding subspace than listing the basis codewords.

A representation of stabilizers that is often useful is as a pair
of binary matrices, frequently written adjacent with a line
between them~\cite{CRSS}.  The first matrix has a $1$ everywhere
the stabilizer has an $X$ or a $Y$, and a $0$ elsewhere; the
second matrix has a $1$ where the stabilizer has a $Y$ or a $Z$.
Multiplying together Pauli operators corresponds to adding the two
rows for both matrices.  Two operators $M$ and $N$ commute iff
their binary vector representations $(a_1 | b_1)$, $(a_2, b_2)$
are orthogonal under a symplectic inner product: $a_1 b_2 + b_1
a_2 = 0$.  For instance, the stabilizer for the five-qubit code
becomes the matrix
\begin{equation}
\left(
\begin{array}{ccccc}
  $1$ & $0$ & $0$ & $1$ & $0$
\\$0$ & $1$ & $0$ & $0$ & $1$
\\$1$ & $0$ & $1$ & $0$ & $0$
\\$0$ & $1$ & $0$ & $1$ & $0$
\end{array}
\right| \left.
\begin{array}{ccccc}
  $0$ & $1$ & $1$ & $0$ & $0$
\\$0$ & $0$ & $1$ & $1$ & $0$
\\$0$ & $0$ & $0$ & $1$ & $1$
\\$1$ & $0$ & $0$ & $0$ & $1$
\end{array}
\right).
\end{equation}

As an example of an application of this representation, let us prove
a fact used above:
\begin{lemma}
Given any stabilizer $S$, there is always at
least one error with any given error syndrome.
\end{lemma}
\begin{proof}
Suppose $S$ has
$a$ generators.  The error syndrome of a Pauli operator $E$ can be
defined as an $a$-component binary vector with the $i$th entry indicating
whether the $i$th generator of $S$ commutes with $E$ (the $i$th bit is $0$) or
anticommutes (the $i$th bit is $1$).  Thus, if $x_i$ is the binary vector representing
the $i$th generator of $S$ and $e$ is the binary vector representing $E$,
then $E$ has error syndrome $v$ iff $x_i \odot e = v_i$, where $\odot$
is the symplectic inner product.  The generators of $S$ give linearly-independent
binary vectors, so we have $a$ independent linear equations in a $2n$-dimensional
binary vector space with $a \leq n$.  By a standard linear algebra theorem, these
equations must always have a non-zero solution.  (In fact, there is a whole
$(2n-a)$-dimensional subspace of solutions.)
\end{proof}

\subsection{Codes over $\GF$}

The CSS construction is very nice in that it allows us to use
the immense existing body of knowledge on classical binary
codes to construct quantum codes.  However, CSS codes cannot
be as efficient as the best stabilizer codes --- for instance,
there is no $[[5,1,3]]$ CSS code.  Instead, if we want to construct
the most general possible stabilizer codes, we should take
advantage of another connection to classical coding theory.

Frequently, classical coding theorists consider not just
binary codes, but codes over larger finite fields.  One of the
simplest is $\GF$, the finite field with four elements.  It is a
field of characteristic $2$, containing the elements $\{0, 1,
\omega, \omega^2\}$.
\begin{equation}
\omega^3 = 1,\ \omega + \omega^2 = 1
\end{equation}
It is also useful to consider two operations on $\GF$.  One is
conjugation, which switches the two roots of the characteristic
polynomial $x^2 + x + 1$:
\begin{align}
\overline{1} & = 1  & \overline{\omega} & = \omega^2
\\ \overline{0} & = 0  & \overline{\omega^2} & = \omega
\end{align}
The other is trace.  $\tr x$ is the trace of the linear operator
``multiplication by $x$'' when $\GF$ is considered as a vector
space over $\mathbb{Z}_2$:
\begin{align}
\tr 0 = \tr 1 & = 0
\\ \tr \omega = \tr \omega^2 & = 1
\end{align}

Stabilizer codes make extensive use of the Pauli group $\Pauli_n$.
We can make a connection between stabilizer codes and codes over
$\GF$ by identifying the four operators $I$, $X$, $Y$, and $Z$
with the four elements of $\GF$, as in
table~\ref{tab:GF4}~\cite{CRSS:GF4}.
\begin{table}
\centering
\begin{tabular}{ll}
Stabilizers & $\GF$
\\ \hline
\\ $I$ & $0$
\\ $Z$ & $1$
\\ $X$ & $\omega$
\\ $Y$ & $\omega^2$
\\ tensor products & vectors
\\
\\ multiplication & addition
\\ $\com{M}{N} = 0$ & $\tr (M \cdot \overline{N}) = 0$
\\ $N(S)$ & dual
\end{tabular}
\caption{Connections between stabilizer codes and codes over
$\GF$.} \label{tab:GF4}
\end{table}

The commutativity constraint in the Pauli group becomes a
symplectic inner product between vectors in $\GF$.  The fact that
the stabilizer is Abelian can be phrased in the language of $\GF$
as the fact that the code must be contained in its dual with
respect to this inner product.  To determine the number of errors
corrected by the code, we must examine vectors which are in the
dual (corresponding to $N(S)$) but not in the code (corresponding
to $S$).

The advantage of making this correspondence is that a great deal
of classical coding theory instantly becomes available.  Many
classical codes over $\GF$ are known, and many of them are
self-dual with respect to the symplectic inner product, so they
define quantum codes.  For instance, the five-qubit code is one
such --- in fact, it is just a Hamming code over $\GF$!  Of
course, mostly classical coding theorists consider {\em linear}
codes (which are closed under addition and scalar multiplication),
whereas in the quantum case we wish to consider the slightly more
general class of \emph{additive} $\GF$ codes (that is, codes which
are closed under addition of elements, but not necessarily scalar
multiplication).

\subsection{Even more quantum error-correcting codes}

There are, of course, many quantum error-correcting codes that are
not stabilizer codes, and a good deal of work has been done on other
sorts of codes.  Usually, you need to assume a certain level of
structure in order to be able to find and work with a code, and there
are a number of ways to ensure that you have sufficient structure
available.

One very fruitful way is to consider codes not over qubits, but over
higher-dimensional registers, \emph{qudits}.  There is a natural
generalization of stabilizer codes to this case~\cite{Knill:qudit},
and a variety of qudit stabilizer codes are known (e.g.,~\cite{AB:Threshold,AK:qudit,KKKS}).
Another route is to relax the stabilizer structure slightly and
look for more efficient qubit codes~\cite{CSSZ}.  One tool that
has garnered interest over the last few years is known as \emph{operator
quantum error correction} or \emph{subsystem codes}~\cite{KLPL,Poulin:subsystem}.
In this case, we ignore certain degrees of freedom in the code,
essentially encoding a state as a linear subspace rather than another
state.  Subsystem codes offer no improvement in the basic error
correction properties I have discussed so far, but do sometimes help
when considering fault tolerance (sections~\ref{sec:ftgates} and
\ref{sec:threshold}).

Another interesting avenue is to study codes which are completely
degenerate.  Such codes are known by various names, most commonly
as \emph{decoherence-free subspaces} (or DFS)~\cite{LW:DFS}.  If all of
the possible errors for a code act as the identity on the code subspace,
then no active correction operation is needed --- no matter what error occurs,
the state remains unchanged.  Usually a DFS is considered for the case
where errors occur continuously in time, in which case the set of
possible errors generates a Lie algebra, and the DFS is then a
degeneracy of the trivial representation of the Lie algebra acting
on the Hilbert space of $n$ qubits.  One advantage of a DFS is that
it continues to function even at very high noise levels, but decoherence-free subspaces have the
disadvantage that a non-trivial DFS only exists for certain very
special noise models (although some, such as collective noise, have
practical significance).  In contrast, a more general QECC can reduce
the effective error rate for a wide variety of types of noise, but
only if the error rate is sufficiently low to begin with.

Of course, one can even go beyond quantum error correction to study
other methods of protecting qubits against noise.  For instance,
in dynamical decoupling~\cite{VKL}, a series of quick operations is performed
which cause the noise to cancel itself out.  Dynamical decoupling only
works when the noise is slowly varying compared to the rate at which
we can perform operations.  It has two advantages: like a DFS, dynamical
decoupling functions well even at relatively high error rates, but unlike
a DFS or QECC, dynamical decoupling does not require any additional qubits.

The problem of eliminating errors from a quantum computer is a difficult
one, and we will want to use every tool that we can bring to bear on the
problem.  Most likely, this will include control techniques like dynamical
decoupling as a first layer of defense, perhaps followed by specialized
error-correcting codes such as a DFS or a phase QECC to handle certain dominant
error types, with a more general quantum error-correcting code as the final
protection to deal with any kinds of errors not eliminated by the first two layers.
However, everything we do --- every qubit we add, every additional gate
we perform --- will have errors in it too, so additional specialized layers
of protection come with a cost, not just in additional overhead, but also
in additional errors that will need to be cleaned up by the final QECC.
It can become a difficult balancing act to judge precisely which protections
are useful and which cause more trouble than they are worth.

\subsection{The logical Pauli group}

The group $N(S)$ has already played an important role in analyzing
a code's ability to correct errors, and it will be nearly as important
later when we discuss fault-tolerant quantum computation.  Therefore,
it is helpful to pause briefly to further consider its structure.

The elements of $N(S)$ are those Pauli operators which commute with
everything in the stabilizer.  Consider how $E \in N(S)$ acts on a
codeword of the stabilizer code.  Let $M \in S$; then
\begin{equation}
M (E \ket{\psi}) = E M \ket{\psi} = E \ket{\psi}.
\end{equation}
This is true $\forall M \in S$, so $E \ket{\psi} \in T(S)$.  That is,
$E$ takes valid codewords to valid codewords.  Now, if $E \in S$ itself,
that is unsurprising: In that case, it takes valid codewords to themselves.
If $E \not\in S$, this cannot be true --- it must take at least one codeword
to a {\em different} codeword.  It is a {\em logical} operation, acting
on the encoded state without interfering with the encoding.\footnote{Incidentally, this
proves that the distance $d$ of a stabilizer code is not accidentally higher
than the distance given in Theorem~\ref{thm:stabilizer}.}  In general, I will
indicate a logical operation by drawing a line over it.  E.g., $X$ is a bit
flip on an individual physical qubit, and $\logical{X}$ is a logical bit flip,
which changes an encoded qubit.

Notice that if $F = EM$, with $M \in S$, then $F \in N(S)$ as well
and $F \ket{\psi} = EM \ket{\psi} = E \ket{\psi}$ for all $\ket{\psi} \in T(S)$.
Thus, two Pauli operators in the same coset of $S$ in $N(S)$ act the
same way, so the different logical operations in $N(S)$ are actually
the elements of $N(S)/S$.  Similarly, note that two Pauli operators
$E$ and $F$ have the same error syndrome iff $E$ and $F$ are in the
same coset of $N(S)$ in $\Pauli_n$.  There is always at least one error
with any given error syndrome, and $\order{\Pauli_n} = 4^{n+1}$, so $\order{N(S)}
= 4 \cdot 2^{n+k}$ and $\order{N(S)/S} = 4^{k+1}$ for an $[[n,k,d]]$ code.

We can in fact identify $N(S)/S$ with the logical Pauli group $\Pauli_k$.  You
can choose any maximal Abelian subgroup $R$ of $N(S)/S$ to represent the logical
$\logical{Z}$ operators (including tensor products of $\logical{Z}$s and $I$s).  The size
of a maximal Abelian subgroup is $4 \cdot 2^k$, since an Abelian subgroup of $N(S)/S$
corresponds to an Abelian subgroup of $\Pauli_n$ which is larger by a factor
$\order{S} = 2^{n-k}$.  By choosing elements of $N(S)/S$ that have various syndromes
with respect to $R$, you can also identify logical $\logical{X}$ operators.  Of
course, in order to make all this work, you need to choose the basis codewords
appropriately.  For instance, the encoded $\ket{\logical{00\dots0}}$ state should
be a $+1$-eigenstate of every element of $R$.

\subsection{The Clifford group}

When working with stabilizer codes, a certain group of quantum gates shows up very
often.  These gates are sufficient to encode and decode stabilizer codes, and play
an important role in the theory of fault-tolerance.  The group is most often known
as the Clifford group (although its relationship to Clifford algebras is tenuous),
and is defined as
\begin{equation}
\Clifford_n = \{ U \in U(2^n) \ | \ UPU^\dagger \in \Pauli_n \ \forall P \in \Pauli_n \}.
\end{equation}
That is, the Clifford group is the normalizer of $\Pauli_n$ in the unitary group
$U(2^n)$.

Besides being important in the theory of stabilizer codes, the Clifford group is
interesting in its own right.  For one thing, it contains some very common quantum
gates.  The Hadamard transform $H$, $\pi/4$ phase rotation $P$, and CNOT gate are
all in the Clifford group:
\begin{equation}
H = \frac{1}{\sqrt{2}} \begin{pmatrix} 1 & 1 \\ 1 & -1 \end{pmatrix}, \quad
P = \begin{pmatrix} 1 & 0 \\ 0 & i \end{pmatrix}, \quad
CNOT = \begin{pmatrix} 1 & 0 & 0 & 0 \\ 0 & 1 & 0 & 0 \\ 0 & 0 & 0 & 1 \\ 0 & 0 & 1 & 0 \end{pmatrix}.
\end{equation}
We can work out how each of these gates acts on the Pauli group by conjugation.
For instance, the Hadamard gate performs the following transformation:
\begin{align}
X &\mapsto Z \\
Z &\mapsto X. \nonumber
\end{align}
There is no need to specify the action of $H$ on $Y$, since conjugation is a
group homomorphism and $Y = iXZ$.  We can therefore immediately determine that
$Y \mapsto iZX = -Y$.  In fact, it turns out that the Clifford group is
generated by $H$, $P$, and CNOT.

In general, to specify a Clifford group operator $U$, it is sufficient to indicate
its action on a generating set for the Pauli group, such as $X$ and $Z$ acting
on each of the $n$ qubits.  This is true because the Pauli group forms a basis
for the $2^n \times 2^n$ matrices, allowing us to learn the action of $U$ on
any projector.  However, there is one remaining ambiguity, since if $U = e^{i\theta} U'$,
then $U$ and $U'$ have the same action by conjugation.  Since this sort of global
phase has no physical significance, however, this is not a very harmful ambiguity.

The Clifford group has a binary matrix representation just like stabilizers do.  Based
on the argument of the last paragraph, we can specify a Clifford group element (up
to global phase) by specifying its action on the binary vectors corresponding to the
Pauli operators.  Since Clifford group elements preserve commutation and anti-commutation,
they correspond to symplectic matrices over the $2n$-dimensional binary vector space.
In fact, we can say more.  The Pauli group $\Pauli_n$ is, by definition, a normal subgroup of
$\Clifford_n$, and because Pauli operators either commute or anticommute, Pauli operators
in $\Clifford_n$ correspond to the identity symplectic matrix.  The center $Z(\Clifford_n)$ consists of
just the diagonal matrices $e^{i\phi} I$, and those also correspond to the identity
symplectic matrix.  Let $\Pauli'_n = Z(\Clifford_n) \Pauli_n$ (that is, the Pauli group, but
with arbitrary phases, not just $\pm 1$, $\pm i$).  Then $\Pauli'_n$ is the kernal of
the map from the Clifford group to the group $Sp(2n,\mathbb{Z}_2)$ of $2n \times 2n$ binary
symplectic matrices.  That is, $\Clifford_n / \Pauli'_n  \cong Sp(2n,\mathbb{Z}_2)$,
which says that if we only care about the action of the Clifford group up to phases,
the Clifford group is effectively just the group of symplectic matrices.

As a consequence of this equivalence, there is an efficient classical simulation of
any circuit of Clifford group operators acting on an initial stabilizer state with final
Pauli measurements~\cite{Gottesman:Heisenberg}.  Even though the overall action of the
circuit is a unitary transformation on a $2^n$-dimensional Hilbert space, each step can be
represented as just a $2n \times 2n$ binary matrix, and we can therefore rapidly
compute the overall circuit action as the product of these matrices.  This result can be
extended to the case where the circuit includes not just unitary Clifford group
gates but also measurements of Pauli operators in the middle of the circuit, with later
gates dependent on the outcome of the measurements.

\section{Fault-tolerant gates}
\label{sec:ftgates}

\subsection{The need for fault tolerance}

There is still a major hurdle before we reach the goal of making quantum
computers resistant to errors.  We must
also understand how to perform operations on a state encoded in a
quantum code without losing the code's protection against errors,
and how to safely perform error correction when the gates used are
themselves noisy.  A protocol which performs these tasks is called
{\em fault tolerant} (FT for short).  Shor presented the first protocols for fault-tolerant
quantum computation~\cite{Shor:FT}, but there have been some
substantial improvements since then.  Now we know that, provided that
the physical error rate per gate and per time step is below some constant
threshold value, it is possible to make the logical quantum computation we wish
to perform arbitrarily close to correct with overhead that is polylogarithmic in
the length of the computation~\cite{AB:Threshold,Kitaev:Threshold,KLZ:Threshold}.

Our goal is to produce protocols which continue to produce the correct answer
even though any individual component of the circuit may fail.  The basic components
which we need to create a universal quantum computer are
\begin{enumerate}
\item {\bf Preparation:} Operations which prepare a new qubit in some standard state.
It is sufficient to have just one type of preparation that prepares a $\ket{0}$ state,
although we will actually use a number of different prepared states.
\item {\bf Quantum Gates:} A universal set of quantum gates.  To have a universal
set, it is sufficient to use the gates $H$, CNOT, and the $\pi/8$ phase rotation
$R_{\pi/8} = \begin{pmatrix} 1 & 0 \\ 0 & e^{i \pi/4} \end{pmatrix}$.  This set of
gates generates a group dense in $U(2^n)$~\cite{BMPRV}.
\item {\bf Measurement:} Measurement of qubits.  It is is sufficient to be able to measure
individual qubits in the standard basis $\ket{0}$, $\ket{1}$.
\item {\bf Wait:} In order to synchronize the operation of gates, we may sometimes need
to have qubits wait around without performing any action on them.
\end{enumerate}
The individual qubits making up our quantum error-correcting code are called {\em physical
qubits}, and each of these actions is a \emph{physical} action (e.g., a physical gate).  Each
instantiation of one of these components is called a \emph{physical location} (or more often
just \emph{location}).  The number of locations in a circuit is then at most the total number of
qubits used times the total number of time steps used.  The number of locations will frequently be
less than the maximum,
as we will often prepare new qubits during the computation and measure qubits, which can then
be discarded, before the computation is completed.  Note that wait steps count as locations,
but that operations on classical data (in particular, measurement results) do not, as we
will assume that classical computation is perfect.  Depending on the precise model we are
using, we may wish to simplify by assuming that modest amounts of classical computation take
no time, but this is not essential.

Any location can fail, including a wait step.  We assume that when a location fails, it
results in an error that can affect all of the qubits
involved in the action.  In the case of preparation, a single-qubit quantum gate,
measurement, or wait, that is just a single qubit.  For a two-qubit quantum gate
such as CNOT, we allow an arbitrary error acting on the two qubits involved in the gate,
including errors which entangle the two qubits.  The actual error should be considered
to be the action of the failed component times the inverse of the desired component
in that location.  Thus, if we wish to perform $Z$, but instead perform $Y$, the
error is $YZ = i X$.  The goal of fault tolerance is to take a quantum circuit which is
designed to work in the absence of errors and modify it to produce a new circuit which
produces the same output as the original circuit, but with the weaker assumption that
the number of failed locations is not too large.  The precise rules for
the probability of locations failing and the type of errors produced when a location
fails will be discussed in section~\ref{sec:threshold}.  I will sometimes refer to
a location with an error in it as a {\em faulty location}.

The biggest obstacle which we must overcome in order to create a fault-tolerant
protocol is that of error propagation.  Even if the gates we perform are themselves
perfect, the action of those gates on the state can alter any errors that have
already occurred and cause them to spread:
\begin{equation}
U E \ket{\psi} = (U E U^\dagger) U \ket{\psi}.
\end{equation}
That is, a pre-existing error $E$ on a state $\ket{\psi}$ followed by a correct gate
$U$ is equivalent to the correct state ($U \ket{\psi}$), but with an error $UEU^\dagger$.
When $U$ is a single-qubit gate, this is not a very serious problem, since the weight
of $E$ does not change, although the exact type of error may now be different.  For instance,
an $X$ error will become a $Z$ error under the action of a Hadamard gate.  The troublesome
case is when $U$ is a two-qubit gate, in which case a single-qubit error $E$ will often
become a two-qubit error.  For instance, notice that CNOT can propagate an $X$ error from
the first qubit to the second, and can propagate $Z$ from the second qubit to the first:
\begin{equation}
CNOT: \quad X \otimes I \mapsto X \otimes X, \quad I \otimes Z \mapsto Z \otimes Z.
\end{equation}
This is a problem because it can increase the weight of an error.  For instance, if
we are using a distance $3$ code, it can handle a single-qubit error, but if we then
perform a CNOT, even if the CNOT itself is perfect, that single-qubit error can become
a two-qubit error, and our distance $3$ code cannot necessarily correct that.
Since we are not going to be able to make a universal quantum computer using only
single-qubit gates, clearly we are going to have be very careful as to how we use
two-qubit gates.

There is, of course, a solution to this problem, which I will discuss in the remainder
of the chapter.  Fault-tolerant circuits will be designed in such a way as to make sure
error propagation does not get out of hand.  Even though errors may spread
somewhat, we can still correct the resulting errors, provided there are not too many
to start with.  Our eventual goal is to produce fault-tolerant versions of all the types
of physical location.  I will refer to each such construction as a \emph{gadget} for the
particular operation.  For instance, we will have fault-tolerant gates for each member of a
universal set of quantum gates.  Each of these gadgets will simulate the behavior of the
corresponding non-fault-tolerant action, but instead of doing so on one or two physical qubits,
it will perform the action on the logical qubits encoded in a quantum error-correcting code.
When we are given a quantum circuit which we would like to perform, we replace each of
the locations in the original circuit with the corresponding fault-tolerant gadget.

Generally, we assume that the original circuit takes no input: all qubits used in it must
be prepared using preparation locations.  This still allows us to perform arbitrary
quantum computations, since we can modify the quantum circuit based on the classical
description of the problem we wish to solve.  (For instance, if we wish to factor a
number $N$, we could tailor the exact quantum circuit to work with $N$.)  Then the
final fault-tolerant measurement gadgets will produce classical information which
should, if the fault-tolerant circuit has done its work properly, give the same outcome
as the original circuit would have if we could have implemented it without error.

\subsection{Definition of fault tolerance}

It is perhaps time to get more precise about exactly what we mean by fault tolerance.
A fault-tolerant gadget should have two basic properties: When the input state to the
gadget does not have too many errors in it, and there are not too many errors on the
physical locations in the gadget, the output state should also not have
too many errors; and, when there are not too many errors in the input state
or during the course of the gadget, the gadget should perform the correct logical operation on
the encoded state.  To define these properties rigorously, we need to first introduce
the notions of an $r$-filter and an ideal decoder~\cite{AGP}.

\begin{defn}
An \emph{$r$-filter} is a projector onto the subspace spanned by all states of the form
$Q \ket{\psi}$, where $\ket{\psi}$ is an arbitrary codeword and $Q$ is a Pauli error of
weight at most $r$.  An \emph{ideal decoder} is a map constructed by taking the input state
and performing a decoding operation (including error correction) consisting of a circuit with
no faulty locations.
\end{defn}
That is, the $r$-filter projects onto states with at most $r$ errors.
Of course, the $r$-filter has no way of knowing what the correct codeword is at this
point of the computation, so even a $0$-filter might project on the wrong state.  The
point is that the only states that can pass through the $r$-filter are those which could
possibly be created from a valid codeword with at most $r$ single-qubit errors.  The
ideal decoder takes the encoded state, corrects any errors, and gives us an unencoded
state.  The ideal decoder gives us a way of talking about the logical state of the quantum computer
at any point during the computation, and the $r$-filter makes precise the notion of a
state having ``at most $r$ errors.''

It is convenient to use a graphical notation
to represent these objects, as follows:
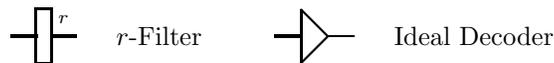
\begin{equation*}
\begin{picture}(200,40)
\thicklines

% r-filter
\put(0,20){\line(1,0){10}}
\put(10,10){\framebox(5,20){}}
\put(15,20){\line(1,0){10}}
\put(18,25){\makebox(0,0)[bl]{$\scriptstyle r$}}
\put(40,12){\makebox(20,16)[l]{\text{$r$-Filter}}}

% ideal decoder

\put(100,20){\line(1,0){10}}
\put(110,10){\line(0,1){20}}
\put(110,30){\line(1,-1){10}}
\put(120,20){\line(-1,-1){10}}
\thinlines
\put(120,20){\line(1,0){10}}
\put(145,12){\makebox(55,16)[l]{\text{Ideal Decoder}}}

\end{picture}
\end{equation*}
%\linebreak
The horizontal lines represent a single block of a QECC, except for the one on the
right end of the ideal decoder symbol, which is a single unencoded qubit.  We will
focus on the case where the code we use is an $[[n,1,2t+1]]$ code.  That is, there is
just one encoded qubit per block, the code can correct $t$ errors, and the thick
horizontal lines in the diagrams represent $n$ qubits.  It is also possible to achieve
fault-tolerance with multiple qubits encoded per block~\cite{Gottesman:FT}, but matters
are somewhat more complicated then.

We are going to need fault-tolerant gadgets representing state preparation, measurement,
and gates.  (The fault-tolerant ``wait'' gadget just consists of having all the
encoded qubits wait.)  In addition, we will need to correct errors during the course of
the computation so that they do not build up to an unacceptable level.  Naturally, our
error correction step also needs to be fault tolerant, since otherwise performing error
correction would have a substantial risk of creating more errors than it fixes.  This
may still happen if the error rate is too high, but at least by designing the error
correction step properly, we have a fighting chance of improving matters by doing
error correction.  We will represent all of these gadgets graphically as well:
\begin{equation*}
\begin{picture}(280,70)

\thicklines
% FT preparation
\put(20,50){\oval(20,20)[l]}
\put(20,40){\line(0,1){20}}
\put(20,50){\line(1,0){10}}
\put(23,55){\makebox(0,0)[bl]{$\scriptstyle s$}}
\put(45,50){\makebox(0,0)[l]{\text{Preparation}}}

% FT measurement
\put(10,20){\line(1,0){10}}
\put(20,20){\oval(20,20)[r]}
\put(20,10){\line(0,1){20}}
\put(31,25){\makebox(0,0)[bl]{$\scriptstyle s$}}
\put(45,20){\makebox(0,0)[l]{\text{Measurement}}}

% FT gate
\put(140,50){\line(1,0){10}}
\put(160,50){\circle{20}}
\put(160,50){\makebox(0,0){$U$}}
\put(170,50){\line(1,0){10}}
\put(171,55){\makebox(0,0)[bl]{$\scriptstyle s$}}
\put(195,50){\makebox(0,0)[l]{\text{Gate $U$}}}

% FT EC
\put(140,20){\line(1,0){10}}
\put(150,10){\framebox(20,20){\text{EC}}}
\put(170,20){\line(1,0){10}}
\put(173,25){\makebox(0,0)[bl]{$\scriptstyle s$}}
\put(195,20){\makebox(0,0)[l]{\text{Error Correction}}}

\end{picture}
\end{equation*}
As before, the thick horizontal lines represent a block of an $[[n,1,2t+1]]$ QECC.
In the case of the encoded gate, if it is a two-qubit logical gate, the
horizontal lines represent {\em two} blocks of the QECC, each containing one logical qubit
involved in the gate.  The $s$ in each diagram represents the maximum number of faulty
locations that may be involved in the circuit represented by the graphic.  For simplicity,
let us restrict attention to cases where the error associated to each fault is a Pauli
operator.  A slight generalization of Theorem~\ref{thm:linear} will allow us to consider
other sorts of errors by looking at linear combinations of diagrams with specific Pauli
errors. If I draw a similar diagram but with thin lines and no indication of the number
of errors, that means the diagram represents an idealized unencoded version of the same
operation.

Now we can say rigorously what it means for these gadgets to be fault tolerant.
The following definitions will involve $t$, the number of errors the code can
correct, and the ideal decoder for the code.  We only need to guarantee the behavior
of the system when the total number of errors involved is less than $t$, since we expect
the constructions to fail no matter what we do when there are more errors than the code can
correct.

\begin{defn}[Fault-Tolerant Measurement]
A measurement gadget is {\em fault tolerant} if it satisfies the following property:
\begin{trivlist}
\item {\bf Meas:} $
\begin{picture}(60,25)
\thicklines

% r-filter
\put(10,5){\line(1,0){10}}
\put(20,-5){\framebox(5,20){}}
\put(25,5){\line(1,0){15}}
\put(28,10){\makebox(0,0)[bl]{$\scriptstyle r$}}

% FT measurement
\put(40,5){\oval(20,20)[r]}
\put(40,-5){\line(0,1){20}}
\put(51,10){\makebox(0,0)[bl]{$\scriptstyle s$}}
\end{picture}
=
\begin{picture}(95,25)
\thicklines

% r-filter
\put(10,5){\line(1,0){10}}
\put(20,-5){\framebox(5,20){}}
\put(25,5){\line(1,0){15}}
\put(28,10){\makebox(0,0)[bl]{$\scriptstyle r$}}

% ideal decoder
\put(40,-5){\line(0,1){20}}
\put(40,15){\line(1,-1){10}}
\put(50,5){\line(-1,-1){10}}
\thinlines
\put(50,5){\line(1,0){15}}

% ideal measurement
\put(65,5){\oval(20,20)[r]}
\put(65,-5){\line(0,1){20}}
\end{picture}
$
when $r+s \leq t$.
\vspace{5pt}
\end{trivlist}
\end{defn}

That is, if the total number of errors in the incoming state and measurement
gadget is at most $t$, then we should get the same result out of the real gadget
as if we had performed ideal decoding on the incoming state and measured the
decoded qubit.  By ``the same result,'' I mean not only that the various measurement
outcomes have the same probability in both cases, but that the remainder of
the computer is left in the same relative state, conditioned on the measurement
outcome, for either diagram.  Really, we are comparing two operations, each of
which transforms a quantum state of the whole computer into a quantum state
for the computer minus one encoded block, plus a classical measurement outcome.
The two operations are the same when the measurement gadget is fault tolerant.

\begin{defn}[Fault-Tolerant Preparation]
A preparation gadget is {\em fault tolerant} if it satisfies the following two
properties:
\begin{trivlist}
\item {\bf Prep A:}
$
\begin{picture}(40,25)
\thicklines

% FT preparation
\put(20,5){\oval(20,20)[l]}
\put(20,-5){\line(0,1){20}}
\put(20,5){\line(1,0){10}}
\put(23,10){\makebox(0,0)[bl]{$\scriptstyle s$}}
\end{picture}
=
\begin{picture}(60,25)
\thicklines

% FT preparation
\put(20,5){\oval(20,20)[l]}
\put(20,-5){\line(0,1){20}}
\put(20,5){\line(1,0){15}}
\put(23,10){\makebox(0,0)[bl]{$\scriptstyle s$}}

% r-filter
\put(35,-5){\framebox(5,20){}}
\put(40,5){\line(1,0){10}}
\put(43,10){\makebox(0,0)[bl]{$\scriptstyle s$}}
\end{picture}
$
when $s \leq t$.
\vspace{5pt}

\item {\bf Prep B:}
$
\begin{picture}(65,25)
\thicklines

% FT preparation
\put(20,5){\oval(20,20)[l]}
\put(20,-5){\line(0,1){20}}
\put(20,5){\line(1,0){15}}
\put(23,10){\makebox(0,0)[bl]{$\scriptstyle s$}}

% ideal decoder
\put(35,-5){\line(0,1){20}}
\put(35,15){\line(1,-1){10}}
\put(45,5){\line(-1,-1){10}}
\thinlines
\put(45,5){\line(1,0){10}}

\end{picture}
=
\begin{picture}(40,25)
\thinlines

% ideal preparation
\put(20,5){\oval(20,20)[l]}
\put(20,-5){\line(0,1){20}}
\put(20,5){\line(1,0){10}}

\end{picture}
$
when $s \leq t$.
\vspace{5pt}
\end{trivlist}

\end{defn}
That is, a fault-tolerant preparation step with $s \leq t$ errors should output a state that is within $s$ errors of a properly encoded state, and that furthermore, the state should decode to the correct state under an ideal decoder.  In the above diagram equation for Prep A, and in many of the equations below, when we have a fault-tolerant gadget on both the left and right side of the equation, assume the faults on both sides are in the same locations and of the same type.

The definitions for a fault-tolerant gate are slightly more complicated, but of much the same form:
\begin{defn}
A gate gadget is {\em fault tolerant} if it satisfies the following two
properties:
\begin{trivlist}
\item {\bf Gate A:}
$
\begin{picture}(80,25)
\thicklines

% r-filter
\put(10,5){\line(1,0){10}}
\put(20,-5){\framebox(5,20){}}
\put(25,5){\line(1,0){15}}
\put(28,10){\makebox(0,0)[bl]{$\scriptstyle r_i$}}

% FT gate
\put(50,5){\circle{20}}
\put(50,5){\makebox(0,0){$U$}}
\put(60,5){\line(1,0){10}}
\put(61,10){\makebox(0,0)[bl]{$\scriptstyle s$}}

\end{picture}
=
\begin{picture}(130,25)
\thicklines

% r-filter
\put(10,5){\line(1,0){10}}
\put(20,-5){\framebox(5,20){}}
\put(25,5){\line(1,0){15}}
\put(28,10){\makebox(0,0)[bl]{$\scriptstyle r_i$}}

% FT gate
\put(50,5){\circle{20}}
\put(50,5){\makebox(0,0){$U$}}
\put(60,5){\line(1,0){15}}
\put(61,10){\makebox(0,0)[bl]{$\scriptstyle s$}}

% (r+s)-filter
\put(75,-5){\framebox(5,20){}}
\put(80,5){\line(1,0){10}}
\put(83,10){\makebox(0,0)[bl]{$\scriptstyle s+ \sum_i r_i$}}

\end{picture}
$
when $s+ \sum_i r_i \leq t$.
\vspace{5pt}

\item {\bf Gate B:}
$
\begin{picture}(105,25)
\thicklines

% r-filter
\put(10,5){\line(1,0){10}}
\put(20,-5){\framebox(5,20){}}
\put(25,5){\line(1,0){15}}
\put(28,10){\makebox(0,0)[bl]{$\scriptstyle r_i$}}

% FT gate
\put(50,5){\circle{20}}
\put(50,5){\makebox(0,0){$U$}}
\put(60,5){\line(1,0){15}}
\put(61,10){\makebox(0,0)[bl]{$\scriptstyle s$}}

% ideal decoder
\put(75,-5){\line(0,1){20}}
\put(75,15){\line(1,-1){10}}
\put(85,5){\line(-1,-1){10}}
\thinlines
\put(85,5){\line(1,0){10}}

\end{picture}
=
\begin{picture}(105,25)
\thicklines

% r-filter
\put(10,5){\line(1,0){10}}
\put(20,-5){\framebox(5,20){}}
\put(25,5){\line(1,0){15}}
\put(28,10){\makebox(0,0)[bl]{$\scriptstyle r_i$}}

% ideal decoder
\put(40,-5){\line(0,1){20}}
\put(40,15){\line(1,-1){10}}
\put(50,5){\line(-1,-1){10}}
\thinlines
\put(50,5){\line(1,0){15}}

% ideal gate
\put(75,5){\circle{20}}
\put(75,5){\makebox(0,0){$U$}}
\put(85,5){\line(1,0){10}}

\end{picture}
$
when $s+ \sum_i r_i \leq t$.
\vspace{5pt}
\end{trivlist}

In all of these diagrams, a separate filter is applied to each input block when $U$ is a multiple-qubit gate.  Input block $i$ gets an $r_i$-filter.  In property Gate A, a separate filter is applied to each output block, but in all cases it is an $s+ \sum_i r_i$-filter.  In property Gate B, an ideal decoder is applied separately to each block.
\end{defn}
Property Gate A says that errors should not propagate too badly: it is OK (and unavoidable) for errors to propagate from one block to another, but they should not spread within a block.  Thus, the final number of errors on the outgoing state of each block should be no more than the total number of errors on the incoming states, plus the number of errors that occurred during the gate gadget.  As before, this only needs to apply when the total number of errors is less than $t$.  Property Gate B says that if there are not too many errors in the incoming blocks and gadget combined, then the fault-tolerant gate gadget should perform the right encoded gate.  Gate B almost says that we can create a commutative diagram with the ideal decoder, the FT gate gadget, and the unencoded ideal gate gadget, but the commutation only need hold when the incoming states have few total errors.

Finally, we must define fault-tolerant error correction:
\begin{defn}
An error correction (EC) gadget is {\em fault tolerant} if it satisfies the following two
properties:
\begin{trivlist}
\item {\bf EC A:}
$
\begin{picture}(60,25)
\thicklines

% FT EC
\put(10,5){\line(1,0){10}}
\put(20,-5){\framebox(20,20){\text{EC}}}
\put(40,5){\line(1,0){10}}
\put(43,10){\makebox(0,0)[bl]{$\scriptstyle s$}}

\end{picture}
=
\begin{picture}(80,25)
\thicklines

% FT EC
\put(10,5){\line(1,0){10}}
\put(20,-5){\framebox(20,20){\text{EC}}}
\put(40,5){\line(1,0){15}}
\put(43,10){\makebox(0,0)[bl]{$\scriptstyle s$}}

% s-filter
\put(55,-5){\framebox(5,20){}}
\put(60,5){\line(1,0){10}}
\put(63,10){\makebox(0,0)[bl]{$\scriptstyle s$}}

\end{picture}
$
when $s \leq t$.
\vspace{5pt}

\item {\bf EC B:}
$
\begin{picture}(105,25)
\thicklines

% r-filter
\put(10,5){\line(1,0){10}}
\put(20,-5){\framebox(5,20){}}
\put(25,5){\line(1,0){15}}
\put(28,10){\makebox(0,0)[bl]{$\scriptstyle r$}}

% FT EC
\put(40,-5){\framebox(20,20){\text{EC}}}
\put(60,5){\line(1,0){15}}
\put(63,10){\makebox(0,0)[bl]{$\scriptstyle s$}}

% ideal decoder
\put(75,-5){\line(0,1){20}}
\put(75,15){\line(1,-1){10}}
\put(85,5){\line(-1,-1){10}}
\thinlines
\put(85,5){\line(1,0){10}}

\end{picture}
=
\begin{picture}(70,25)
\thicklines

% r-filter
\put(10,5){\line(1,0){10}}
\put(20,-5){\framebox(5,20){}}
\put(25,5){\line(1,0){15}}
\put(28,10){\makebox(0,0)[bl]{$\scriptstyle r$}}

% ideal decoder
\put(40,-5){\line(0,1){20}}
\put(40,15){\line(1,-1){10}}
\put(50,5){\line(-1,-1){10}}
\thinlines
\put(50,5){\line(1,0){10}}

\end{picture}
$
when $r+s \leq t$.
\vspace{5pt}

\end{trivlist}

\end{defn}
That is, after an error correction step with at most $s$ faulty locations, the state is at most $s$ errors away from some encoded state.  Note that this must apply no matter how many errors were in the incoming state.  This does not necessarily mean those errors were dealt with properly, only that the final state is near a codeword.  It might be the \emph{wrong} codeword, but it is still a valid codeword.  Property EC B does say that if the total number of incoming errors and errors during the FT EC step is less than $t$, the state has been corrected, in the sense that the logical state after the EC step is the same as the logical state before it.

\subsection{Transversal gates}

Now we must try to find constructions that fulfill these definitions.  Let us start with gate gadgets.  Indeed, we have already seen a construction of fault-tolerant gates: Recall that for an $[[n,k,d]]$ stabilizer code with stabilizer $S$, $N(S)/S \cong \Pauli_k$, the logical Pauli group on the $k$ encoded qubits.  Thus, in the absence of any errors, we can perform a logical $\logical{Z}$, for instance, by choosing a representative $Q$ of the appropriate coset in $N(S)/S$.  $Q \in \Pauli_n$, so to perform it, we simply need to perform some Pauli matrix (or the identity) on each of the $n$ physical qubits in the code.  Observe that this construction satisfies properties Gate A and Gate B: Since we are performing single-qubit gates, there is no opportunity for errors to propagate to different qubits, so property Gate A is clearly satisfied for any number of errors (even if it is greater than $t$).  Property Gate B follows from this fact as well (although in this case, we really do need to specify that the number of errors is at most $t$).  If you want to prove more formally that these properties are satisfied, the key step is to note that we can rearrange the errors to all come after the logical gate.  Moving errors through the Pauli operator $Q$ may change their identity --- we conjugate by $Q$ --- but does not change their weight.

Of course, the reason this is true is that $Q$ is a tensor product of single-qubit gates.  For many codes, we can find additional gates that can be performed this way.  For instance, for the $7$-qubit code, you can check that performing the Hadamard on all $7$ qubits, $H^{\otimes 7}$, performs the logical Hadamard $\logical{H}$.  This construction of $\logical{H}$ is automatically fault-tolerant, again because conjugating an error by $\logical{H}$ does not change the weight of the error.  Similarly, for the $7$-qubit code, one can fault-tolerantly perform $\overline{P}$ (the $\pi/4$ rotation) by performing $P^\dagger$ on each of the $7$ physical qubits in the code~\cite{Shor:FT}.

If we wish to generalize to multiple-qubit gates --- and we need to do that somehow to get universal quantum computation --- we can no longer assume that conjugating by $U$ will leave the weight of an error unchanged.  However, if we make sure that any increase in the number of errors is spread out between multiple blocks of the code, we can ensure that each block is not overwhelmed with errors.  In particular, suppose we consider a gate (which may act on $m$ blocks of the code) which is constructed as a tensor product $U = \bigotimes U_i$, where $U_i$ acts on the $i$th qubit of each block.  The $U_i$s can do whatever they like to the $m$ qubits they act on; we don't even make the constraint that all the $U_i$s be the same.  A gate constructed this way is called a {\em transversal} gate, and it will automatically be fault tolerant.  When $U_i$ conjugates any number of errors on the $m$ $i$th qubits, we can get an error that acts only on those $m$ qubits. In particular, $U_i$ will only ever propagate a single error into at most one qubit per block.  Therefore, properties Gate A and Gate B will also hold for transversal gates.

Note that we insist in the definition of transversal that the $i$th qubit of one block only interacts with the $i$th qubit of another block, and not some different qubit in the second block.  If this were not true, the gate would still be fault tolerant provided each qubit in one block interacted with only a single qubit in a different block.  However, such a construction could cause problems if we attempted to put two of them together.  For instance, if we have a gate that interacts qubit $i$ of block $1$ with qubit $j \neq i$ of block $2$, and then follow it by a traditional transversal gate, an initial error on qubit $i$ of block $1$ could propagate to qubit $j$ of block $2$ and then back to qubit $j$ of block $1$, leaving us with two errors in the first block.  The product of two transversal gates is again transversal, and thus fault-tolerant, but this would not be true if we changed the definition of transversal to allow permutations of the qubits in a block.

As an example of a transversal two-qubit gate, consider again the $7$-qubit code.  Performing $\text{CNOT}^{\otimes 7}$ --- a transversal CNOT between corresponding qubits of the two blocks --- implements the logical $\logical{\text{CNOT}}$~\cite{Shor:FT}.  We sometimes use the terminology ``transversal $U$'' for the transversal gate $U^{\otimes n}$.  Even though transversal does not necessarily mean that all the tensor components are the same, the most common examples have this property, so the phrase is often used as a shorthand.

In fact, the transversal CNOT performs the logical $\logical{\text{CNOT}}$ for any CSS code, not just the $7$-qubit code.  This fact is one reason that CSS codes are particularly favorable for fault tolerance.  (We shall see another in the following section.)  The $7$-qubit code is {\em particularly} good. As I noted above, for the $7$-qubit code, we have transversal implementations of $\logical{H}$, $\logical{P}$, and $\logical{\text{CNOT}}$.  Since products of transversal gates are again transversal, this implies that the whole logical Clifford group can be implemented transversally as well.  However, the Clifford group is not universal for quantum computation (it can even be efficiently simulated classically), and unfortunately, there are no other transversal logical gates for the $7$-qubit code, so we will have to resort to another type of construction to complete a universal set of logical gates.  One might hope to avoid this by finding a better code, but in fact, no code allows a universal set of transversal gates~\cite{EK}.

\subsection{Fault-tolerant error correction and measurement}

We will return to the task of completing the universal set of logical gates in a little bit, but first let us discuss another part of a fault-tolerant protocol, fault-tolerant error correction.  Before we do that, we should first consider precisely how we do {\em non}-fault-tolerant error correction.

\begin{figure}
\begin{centering}
\begin{picture}(140,100)

\put(40,90){\line(1,0){34}}
\put(0,90){\makebox(30,0){$\ket{0} + \ket{1}$}}
\put(74,82){\framebox(16,16){$H$}}
\put(90,90){\line(1,0){10}}
\put(100,90){\line(1,1){10}}
\put(100,90){\line(1,-1){10}}
\put(115,90){\makebox(0,0)[l]{Eigenvalue}}

\multiput(20,10)(0,20){3}{\line(1,0){30}}
\put(60,90){\circle*{6}}
\put(50,2){\framebox(20,56){$U$}}
\put(60,90){\line(0,-1){32}}
\multiput(70,10)(0,20){3}{\line(1,0){60}}
\end{picture}

\caption{A non-fault-tolerant implementation of the measurement of $U$, which has eigenvalues $\pm 1$.}
\label{fig:notFTmeasure}
\end{centering}
\end{figure}
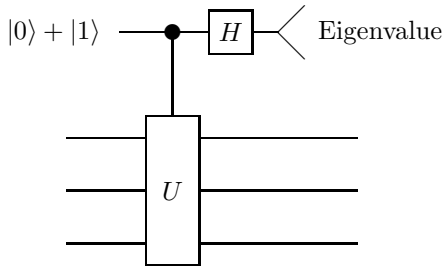

When we have an $[[n,k,d]]$ stabilizer code, to do error correction, we wish to measure the eigenvalue of each of the $n-k$ generators of the stabilizer.  Each eigenvalue tells us one bit of the error syndrome.  It is straightforward to measure the eigenvalue of any unitary operator $U$ using a standard phase kickback trick (see figure~\ref{fig:notFTmeasure}).  Add a single ancilla qubit in the state $\ket{0} + \ket{1}$, and perform the controlled-$U$ from the ancilla qubit to the data for which we wish to measure $U$. Then if the data is in an eigenstate of $U$ with eigenvalue $e^{i\phi}$, the ancilla will remain unentangled with the data, but is now in the state $\ket{0} + e^{i\phi} \ket{1}$.  (If the data is not in an eigenstate, than the ancilla becomes entangled with it, decohering the data in the eigenbasis of $U$.)  In the case where the eigenvalues of $U$ are $\pm 1$, as for $U \in \Pauli_n$, we can just measure $\phi$ by performing a Hadamard transform on the ancilla qubit and measuring it.

However, this construction allows for runaway error propagation, even when $U$ is as simple as a multiple-qubit Pauli operator.  The controlled-$U$ is implemented as a series of controlled-$Q_i$ gates, with $Q_i \in \Pauli_1$ acting on the $i$th data qubit, but each controlled-$Q_i$ gate is capable of propagating errors in either direction between the ancilla qubit and the $i$th data qubit.  Therefore, a single error early on in this construction could easily propagate to many qubits in the data block.

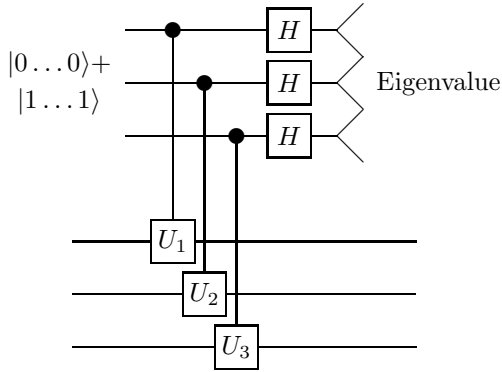
\begin{figure}
\begin{centering}
\begin{picture}(160,140)

\multiput(40,90)(0,20){3}{\line(1,0){54}}
\put(0,117){\makebox(30,0){$\ket{0\dots0} +$}}
\put(0,103){\makebox(30,0){$\ket{1\dots1}$}}
\multiput(94,82)(0,20){3}{\framebox(16,16){$H$}}
\multiput(110,90)(0,20){3}{\line(1,0){10}}
\multiput(120,90)(0,20){3}{\line(1,1){10}}
\multiput(120,90)(0,20){3}{\line(1,-1){10}}
\put(135,110){\makebox(0,0)[l]{Eigenvalue}}

\put(20,10){\line(1,0){54}}
\put(20,30){\line(1,0){42}}
\put(20,50){\line(1,0){30}}

\put(90,10){\line(1,0){60}}
\put(78,30){\line(1,0){72}}
\put(66,50){\line(1,0){84}}

\multiput(58,130)(12,-20){3}{\circle*{6}}
\put(74,2){\framebox(16,16){$U_3$}}
\put(62,22){\framebox(16,16){$U_2$}}
\put(50,42){\framebox(16,16){$U_1$}}
\multiput(58,130)(12,-20){3}{\line(0,-1){72}}

\end{picture}
\caption{A component of a fault-tolerant implementation of the measurement of $U = \bigotimes U_i$, which has eigenvalues $\pm 1$.}
\label{fig:shormeasure}
\end{centering}
\end{figure}

To avoid this, we would like to spread out the ancilla qubit to make the controlled-$U$ gate more like a transversal gate.  However, we want either all the $Q_i$s to act or none of them.  We can achieve this by using an ancilla in the $n$-qubit ``cat'' state $\ket{00\dots0} + \ket{11\dots1}$, as in figure~\ref{fig:shormeasure}.  After interacting with the data, we would like to distinguish the states $\ket{00\dots0} \pm \ket{11\dots1}$.  The most straightforward way to do this is to note that
\begin{align}
H^{\otimes n} (\ket{00\dots0} + \ket{11\dots1}) &= \sum_{\wt(x)=\text{even}} \ket{x} \\
H^{\otimes n} (\ket{00\dots0} - \ket{11\dots1}) &= \sum_{\wt(x)=\text{odd}} \ket{x}.
\end{align}
Thus, by measuring each qubit of the ancilla in the Hadamard basis and taking the parity, we can learn the eigenvalue of $U$.
Now the $i$th qubit of the data only interacts with the $i$th qubit of the ancilla, so there is no chance of catastrophic error propagation.  Errors can only propagate between a single qubit of the ancilla and a single qubit of the data.  For simplicity, I described the cat state as an $n$-qubit state, but of course, it only need be as large as the weight of $U$, since any additional qubits in the ancilla will not interact with the data at all.

Still, this construction does not yet give us a fault-tolerant error correction gadget.  There are two remaining problems.  First, we have not specified how to create the ancilla cat state yet.  Second, if even a single qubit of the ancilla has an error, the measurement result could have the wrong parity, giving us an incorrect syndrome bit.  The solution to the second problem is tedious but straightforward: after measuring every bit of the error syndrome to get a candidate error syndrome, we repeat the process.  If we repeat enough times, and the number of errors in the course of process is not too large, we can eventually be confident in the outcome, as a single faulty location can only cause a single measurement outcome to be incorrect.  Actually, there are some additional complications due to the possibility of errors occurring in the data qubits --- which can cause the true error syndrome to change --- in the middle of the correction process, but these too can be handled by sufficient repetition of the measurement.

To create a cat state, we must be somewhat careful.  The obvious circuits to do it are not fault-tolerant.  For instance, we could put a single qubit in the state $\ket{0} + \ket{1}$, and then perform CNOTs to the other $n-1$ qubits, which are initially in the $\ket{0}$ state.  However, an error halfway through this procedure could give us a state such as $\ket{0011} + \ket{1100}$, which effectively has two bit flip errors on it.  When we interact with the data as in figure~\ref{fig:shormeasure}, the two bit flip errors will propagate into the data block, resulting in a state with $2$ errors in it.  To avoid this, after creating the cat state, we must verify it.  One way to do so is to take pairs of qubits from the cat state and CNOT them both to an additional ancilla qubit which is initialized to $\ket{0}$.  Then we measure the additional ancilla, and if it is $\ket{1}$, we know that the two qubits being tested are different, and we discard the cat state and try again.  Even though this verification procedure is not transversal, it is still going to allow us to fault-tolerantly verify the state due the nature of error propagation in a CNOT.  The ancilla qubit is the target of both CNOT operations, which means only phase errors can propagate from it into the cat state.  A single phase error can already ruin our cat state, giving us the wrong syndrome bit as the outcome --- that is why we must repeat the measurement --- so two phase errors are no worse. If we do sufficient verification steps, we can be confident that either the ancilla cat state is correct, with possibly $s$ errors on it, or there were more than $s$ faulty locations in total during the preparation and verification step for the cat state.  That is enough to ensure that the error correction procedure, taken as a whole, is fault tolerant.

To summarize, we must first create many cat states.  We do this via some non-fault-tolerant procedure followed by verifying pairs of qubits within the cat state to see if they are the same.  We use each cat state to measure one bit of the error syndrome, and repeat the measurement of the full syndrome a number of times.  In the end, we deduce a consensus error syndrome and from it an error which we believe occurred on the data, and then correct that error.  The above procedure is known as Shor error correction~\cite{Shor:FT}, and it works for any stabilizer code.  A similar procedure can be used for measuring a logical Pauli operation on a stabilizer code, although in that case, we must be careful to perform error correction before repeating our measurement attempt, as an uncorrected error on the data can cause the measurement outcome to be incorrect even if the measurement procedure itself is completely free of errors.  I have been somewhat cavalier about some of the details above partially because they are complicated and partly because Shor error correction is rather laborious, requiring many extra gates and a low physical error rate to work well.  There are some much more efficient schemes for fault-tolerant error correction available, so Shor error correction is rarely used.

The first such improved scheme is Steane error correction~\cite{Steane:EC}.  Steane error correction works only on CSS codes.  (There is a more complicated version that works on general stabilizer codes, but it is not usually used; the version in Steane's paper has an error.)  Recall that the codewords of a CSS code are of the form $\sum_{w \in C_2^\perp} \ket{u + w}$ where $C_2$ is a classical linear code, and $u \in C_1$.  $u$ tells us the logical codeword encoded by this state.  If we were to measure every qubit of a CSS code, in the absence of error, we would get the classical result $u+w$ for some random $w \in C_2^\perp$.  Since $C_2^\perp \subseteq C_1$, $u+w \in C_1$, but we can deduce $u$ by seeing which coset from $C_1/C_2^\perp$ $u+w$ falls into.  If there are some errors on the state, phase errors in the original state do not cause errors in the measurement outcome; however, a bit flip error on qubit $i$ produces a bit flip error on the corresponding classical bit $i$.  Thus the classical outcome is $u+w+e$, where $e$ is a vector representing the locations of the errors.  But $C_1$ is a classical error-correcting code, so by applying the classical decoding procedure, we can deduce $e$, $u$, and $w$ if there are not too many errors.

If there are some faults in the physical measurement locations, then the outcome becomes $u+w+e+f$, where $e$ represents the pre-existing bit flip errors before the measurement and $f$ represents the errors caused by faulty measurement locations.  Note, however, that measurement is performed transversally, so a single failed measurement only affects a single bit of the output; that is, $\wt(f) \leq s$ when there are $s$ faulty measurement locations.  In this case, provided $\wt (e+f) \leq t_1$ (assume the classical code $C_1$ corrects $t_1$ errors), we can deduce $u$, $w$, and $e+f$; however, we cannot distinguish $e$ and $f$.  This is annoying, but not particularly harmful if we only wish to measure the state.  In that case, $u$ tells us the outcome of the measurement, and as long as we learn that, we are OK.  This gives us a fault-tolerant measurement procedure for CSS codes.

Of course, we have not achieved fault-tolerant error correction yet.  Measuring the qubits directly does tell us about bit flip errors, but only at the cost of destroying the code block.  Clearly that is not desirable.   We will also need to learn about any phase flip errors in the state.  To do so, we will again introduce some ancilla qubits.  I noted above that transversal CNOT applies the logical $\logical{\text{CNOT}}$ for any CSS code.  Therefore, let us create an ancilla block in a codeword of the CSS code we are using.  Then do a transversal CNOT from the data block to the ancilla block.  Afterwards, both the data block and the ancilla block are still in codeword states.  Furthermore, any bit flip errors in the data were propagated forward along the CNOTs to the corresponding locations in the ancilla block.  Now if we measure every qubit of the ancilla block, as described above, we learn $e+f$ for the ancilla block.  That will be a combination of the bit flip errors on the data block, pre-existing bit flip errors on the ancilla, and bit flip errors that occurred during transversal CNOT or measurement.  There are clearly a lot of extraneous errors to worry about, but at least we have learned something about the errors in the data block.

Still, we need to be careful.  We used a transversal CNOT to copy the errors from the data block to the ancilla, but we don't actually want to perform the logical $\logical{\text{CNOT}}$ gate.  Error correction should leave the encoded data unchanged.  Therefore, we should start the ancilla not in just any codeword state, but in the encoded state $\ket{\logical{0}} + \ket{\logical{1}}$, an eigenstate of $\logical{\text{CNOT}}$ (when it is in the target block of the CNOT).  Thus, the encoded state of the data does not change when we perform the CNOT.  This also means that measuring $u$ for the ancilla tells us nothing about the state of the encoded data; indeed, $u$ will be random, just like $w$.  We have to also be careful about error propagation: while bit flip errors are propagating from the data block into the ancilla block, phase errors are propagating from the ancilla block into the data block.  Therefore, we must be careful that the procedure we use to create the ancilla block does not result in too many errors.  Since the ancilla block is just an encoded $\ket{\logical{0}} + \ket{\logical{1}}$ state, the problem of creating such a state is identical to the problem of creating a fault-tolerant preparation gadget, so I will defer discussion of how to do this until section~\ref{sec:prep}.  For now, just assume that we have such a method which, when the encoding circuit has at most $s$ faulty locations, creates the correct state with at most $s$ errors on it.

\begin{figure}
\begin{centering}
\begin{picture}(150,70)

\thicklines

\put(0,60){\makebox(30,0){data}}
\put(30,60){\line(1,0){120}}

\put(40,30){\line(1,0){50}}
\put(0,30){\makebox(40,0){$\ket{\logical{0}} + \ket{\logical{1}}$}}
\put(40,10){\line(1,0){22}}
\put(20,10){\makebox(10,0){$\ket{\logical{0}}$}}

\put(50,10){\circle*{6}}
\put(50,60){\circle{12}}
\put(50,10){\line(0,1){56}}

\put(70,60){\circle*{6}}
\put(70,30){\circle{12}}
\put(70,60){\line(0,-1){36}}

\put(62,2){\framebox(16,16){$H$}}
\put(78,10){\line(1,0){12}}

\multiput(90,10)(0,20){2}{\line(1,1){10}}
\multiput(90,10)(0,20){2}{\line(1,-1){10}}
\put(105,20){\makebox(0,0)[l]{Error syndrome}}

\end{picture}
\caption{Steane Error Correction.  Each horizontal line represents a full $n$-qubit block of the code, and each gate or measurement represents a transversal implementation of that operation.}
\label{fig:Steane}
\end{centering}
\end{figure}
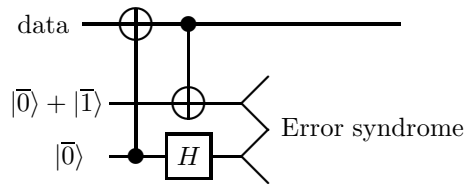

To correct bit flip errors on a CSS code, we thus create an ancilla encoded in the same code in the state $\ket{\logical{0}} + \ket{\logical{1}}$, and perform transversal CNOT from the data block to the code block.  Then we measure all of the qubits of the ancilla block, and treat the result as a classical codeword for the code $C_1$.  We deduce the locations of the errors, and correct them in the data block.  We can follow almost the same procedure to correct phase errors: Instead, we create an ancilla block in the state $\ket{\logical{0}}$ and perform transversal CNOT with the ancilla block as control and the data block as target.  This copies phase errors from the data block to the ancilla, and so we measure each ancilla qubit in the Hadamard-rotated basis.  In that basis, the code is a superposition of codewords from $C_2$, so we treat the measurement output as a classical codeword for $C_2$ with some errors, and deduce those errors via the classical decoding procedure for $C_2$.  Then we correct phase errors in the resulting locations in the data block.  The whole procedure is summarized in figure~\ref{fig:Steane}.

Now let us show that Steane EC satisfies properties EC A and EC B.  Assume that the incoming data block has errors in locations summarized by a vector $e$, with $e$ composed of $e_P$, phase errors, and $e_B$, bit flip errors.  In each of these vectors, $1$ in the $i$th component of the vector indicates an error acting on the $i$th qubit of the block.  $e_P$ and $e_B$ can have $1$s in the same location, in which case there is a $Y$ error at that location.  We can assume that all errors in the incoming block are Pauli errors by linearity (a la Theorem~\ref{thm:linear}).

Suppose the phase ancilla, after its preparation is complete, has errors whose locations can be summarized by a binary vector $f_P$, and the bit flip ancilla has errors that can be summarized by a binary vector $f_B$.  Let us summarize the faults in CNOT gate locations by $g_P$ and $g_B$.  (Note that each such faulty gate location can affect both the corresponding data qubit and ancilla qubit, and not necessarily with the same sort of error.)  Finally, assume that during the Hadamard and measurement of the phase ancilla block, there are faults whose locations are summarized together by a vector $h_P$, and during the measurement of the bit flip ancilla block, there are faults summarized by $h_B$.  For $h_P$ and $h_B$, we will only count errors that show up in the measurement result, since there is no possibility of these errors propagating into the data block except via the measurement result.  For the other sorts of errors (other than those in the incoming data block), we will make no restrictions as to the types of errors.  We will ignore errors that occur during the final correction step where we fix the error deduced from the syndrome measurement, since the correction step itself can be done transversally.  Errors that occur then count, of course, but each error during the final correction step can only cause one qubit to go wrong, so they can never cause EC A or EC B to fail.

To prove EC A, we assume that $s = \wt(f_P) + \wt(f_B) + \wt(g_P) + \wt(g_B) + \wt(h_P) + \wt(h_B) \leq t$, and we wish to show that the final state after correcting the measured error is at most $s$ errors away from a valid codeword of some sort.  In fact, I claim the following:
\begin{claim}
The final state will have errors only on qubits from $f_P \vee f_B \vee g_P \vee g_B \vee h_P \vee h_B$, where $\vee$ represents the bitwise OR.
\end{claim}
That is, the only qubits in the final state that have errors will be in a location corresponding to one of the faults that occurred during the EC circuit, and there are at most $s$ of those.  By ``corresponding location,'' I mean that if a fault occurs in the $i$th qubit of one of the ancilla blocks, there could be a fault in the $i$th qubit of the data block.

\begin{proof}
In the following proof, the $f_x^{(i)}$, $g_x^{(i)}$, and $h_x^{(i)}$ ($x = P, B$) will be vectors which indicate subsets of $f_x$, $g_x$, and $h_x$.  The content of the $(i)$ superscripts has no particular meaning except to label potentially different subsets.

The error measured by the phase part of the error correction will be $e_P + f_P^{(1)} + g_P^{(1)} + h_P^{(1)}$.  However, note that, since we are making no restriction on $\wt(e_P)$, it is possible that $\wt(e_P + f_P^{(1)} + g_P^{(1)} + h_P^{(1)}) > t_2$, where the phase code $C_2$ corrects $t_2$ errors.  In that case, the classical decoding procedure for $C_2$ will instead deduce some error $d$ which might be different from the actual accumulated error.  Still, $d$ must have the same syndrome as $e_P + f_P^{(1)} + g_P^{(1)} + h_P^{(1)}$, so $v = d + e_P + f_P^{(1)} + g_P^{(1)} + h_P^{(1)} \in C_2$.  Since the $X$ generators of the CSS code are given by vectors of $C_2^\perp$, it follows that the operator $Z_v$ with $Z$s on all the qubits indicated by $v \in C_2$ commutes with every $X$ generator, and therefore $Z_v \in N(S)$, and it maps any logical codeword to another (potentially different) logical codeword.

After interacting with the two ancilla blocks, but before we correct any errors, the data will have phase errors on qubits indicated by the set $e_P + g_P^{(2)} + g_B^{(2)} + f_B^{(2)}$.  When we apply phase error correction to fix the errors on the qubits indicated by $d$, we therefore get the net phase error $d + e_P + g_P^{(2)} + g_B^{(2)} + f_B^{(2)}$.  Now, we know that $d + e_P = v + f_P^{(1)} + g_P^{(1)} + h_P^{(1)}$, so we can rewrite the net phase error as $v + f_P^{(1)} + f_B^{(2)} + g_P^{(1)} + g_P^{(2)} + g_B^{(2)} + h_P^{(1)}$.  But $Z_v \in N(S)$, so applying a phase to all the qubits of $v$ only takes us to another codeword.  Therefore, the final state, post-error correction, is equal to some codeword with a phase error $f_P^{(1)} + f_B^{(2)} + g_P^{(1)} + g_P^{(2)} + g_B^{(2)} + h_P^{(1)}$ (plus possibly some bit flip errors).  Note that this is a subset of $f_P \vee f_B \vee g_P \vee g_B \vee h_P \vee h_B$, showing that the phase errors on the final codeword satisfy the claim.

The proof that the final bit flip errors are a subset of $f_P \vee f_B \vee g_P \vee g_B \vee h_P \vee h_B$ is similar.
\end{proof}

Proving the property EC B is more straightforward, since we can now assume that $s + \wt(e) \leq t$.  Again, the final errors will be a subset of $f_P \vee f_B \vee g_P \vee g_B \vee h_P \vee h_B$, but now there is no shift by an element of $N(S)$, so an ideal decoder after the EC step will produce the same result as an ideal decoder applied before the EC step, when there is just the error $e$.

\begin{figure}
\begin{centering}
\begin{picture}(200,60)

\thicklines

\put(0,50){\makebox(30,0){data}}
\put(30,50){\line(1,0){62}}

\put(40,30){\line(1,0){80}}
\put(0,30){\makebox(40,0){$\ket{\logical{0}} + \ket{\logical{1}}$}}
\put(40,10){\line(1,0){102}}
\put(20,10){\makebox(10,0){$\ket{\logical{0}}$}}

\put(60,30){\circle*{6}}
\put(60,10){\circle{12}}
\put(60,30){\line(0,-1){26}}

\put(80,50){\circle*{6}}
\put(80,30){\circle{12}}
\put(80,50){\line(0,-1){26}}

\put(92,42){\framebox(16,16){$H$}}
\put(108,50){\line(1,0){12}}

\multiput(120,30)(0,20){2}{\line(1,1){10}}
\multiput(120,30)(0,20){2}{\line(1,-1){10}}
\put(135,55){\makebox(0,0)[l]{2 cbits plus}}
\put(135,44){\makebox(0,0)[l]{error syndrome}}

\put(150,38){\vector(0,-1){16}}
\put(142,2){\framebox(16,16){$Q$}}
\put(158,10){\line(1,0){22}}

\end{picture}
\caption{Knill EC.  Each horizontal line represents a full $n$-qubit block of the code, and each gate or measurement represents a transversal implementation of that operation.  $Q$ is a Pauli operator and contains a correction for both the teleportation outcome and error correction.}
\label{fig:KnillEC}
\end{centering}
\end{figure}
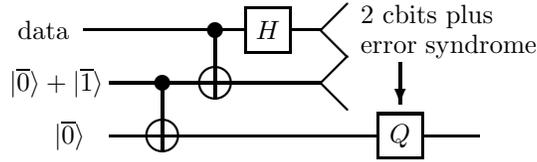

Another useful method of fault-tolerant error correction is due to Knill~\cite{Knill:EC}.  Knill error correction is at some level similar to Steane error correction, in that it uses an ancilla state which is encoded using the same code as the data block.  However, Knill EC works for any stabilizer code.  The basic idea is to perform quantum teleportation, moving the encoded state into a different block of the code.  Because of the encoding, the Bell measurement used in quantum teleportation gains more information than is needed for teleportation, and the extra information tells us the error syndrome of the combined errors on the data and ancilla block.  Knill error correction is pictured in figure~\ref{fig:KnillEC}.
Two blocks are measured in the course of teleportation.  One is basically used to reconstruct the $X$s in the stabilizer and one to reconstruct the $Z$s.  I will not go into the details of how Knill EC works here --- see~\cite{Knill:EC} instead.  Knill EC can easily be modified to allow fault-tolerant measurement as well.  To do that, we substitute an ancilla in the state $\ket{\logical{0}}$ for the encoded EPR pair in the error correction circuit.  The encoded Bell measurement then still tells us the error syndrome, but instead of two random classical bits for the teleportation, it tells us the outcome of a logical measurement on the data block in the standard basis.

Steane and Knill EC both work on the principle of moving as much work as possible in the EC step into the creation of a particular ancilla state.  There are two advantages to doing so~\cite{Knill:EC,Reichardt:ancilla,Steane:EC}.  First, it means we do not have to do much work directly on the data, which means the qubits don't have to sit around waiting for the EC step to finish and don't accumulate much additional error during that time.  Second, because the ancillas are created in {\em known} states, we can put a lot of effort into verifying that they are correct, which we could not do directly for the unknown data state.

\subsection{Gate teleportation and universal fault-tolerant quantum computation}

Now let us return to the task of creating a universal set of fault-tolerant gates.  Since it is not possible to do this transversally, we need to add a new component, and that component will be measurement.  The basic strategy will be similar to that for Steane or Knill EC: we will put a lot of effort into creating an appropriate ancilla state, and then quickly interact it with the data, the result being an implementation of some particular gate.  Not every gate can be directly performed this way, but some non-Clifford group gates can be, and combined with the Clifford group gates, that is enough to give us a universal set of quantum gates.

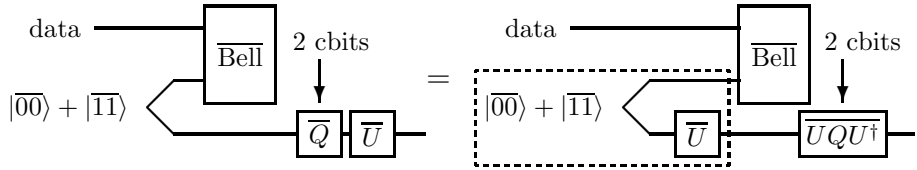
\begin{figure}
\begin{centering}
\begin{picture}(350,60)

\thicklines

\put(0,50){\makebox(30,0){data}}
\put(30,50){\line(1,0){42}}

\put(60,30){\line(1,0){12}}
\put(0,20){\makebox(40,0){$\ket{\logical{00}} + \ket{\logical{11}}$}}
\put(60,10){\line(1,0){47}}
\put(50,20){\line(1,1){10}}
\put(50,20){\line(1,-1){10}}

\put(72,22){\framebox(26,36){$\logical{\text{Bell}}$}}

\put(105,45){\makebox(0,0)[l]{2 cbits}}

\put(115,38){\vector(0,-1){16}}
\put(107,2){\framebox(16,16){$\logical{Q}$}}
\put(123,10){\line(1,0){4}}
\put(127,2){\framebox(16,16){$\logical{U}$}}
\put(143,10){\line(1,0){12}}

\put(160,30){\makebox(0,0){{\Large $=$}}}

\put(180,50){\makebox(30,0){data}}
\put(210,50){\line(1,0){64}}

\put(240,30){\line(1,0){34}}
\put(180,20){\makebox(40,0){$\ket{\logical{00}} + \ket{\logical{11}}$}}
\put(240,10){\line(1,0){10}}
\put(230,20){\line(1,1){10}}
\put(230,20){\line(1,-1){10}}

\put(274,22){\framebox(26,36){$\logical{\text{Bell}}$}}

\put(306,45){\makebox(0,0)[l]{2 cbits}}

\put(250,2){\framebox(16,16){$\logical{U}$}}
\put(266,10){\line(1,0){31}}

\put(313,38){\vector(0,-1){16}}
\put(297,2){\framebox(32,16){$\logical{UQU^\dagger}$}}
\put(329,10){\line(1,0){11}}

\put(174,-2){\dashbox{2}(96,36){}}

\end{picture}
\caption{Gate teleportation of $\logical{U}$.  The process of teleporting the state followed by $\logical{U}$ is the same as teleporting the state through a special ancilla with an appropriately modified correction operation.}
\label{fig:gateteleport}
\end{centering}
\end{figure}

The construction can be described in terms of quantum teleportation, again on the encoded states.  Let us begin by not worrying about fault-tolerance, and simply attempt to perform some gate $\logical{U}$ on an encoded state.  We will consider the case of a single-qubit gate $U$ first.  Suppose we were to perform quantum teleportation and then follow it, somehow, by an implementation of $\logical{U}$.  Then on input state $\ket{\psi}$, the overall output state would of course be $\logical{U} \ket{\psi}$.  Now imagine that Bob, who controls the output block, gets impatient waiting for Alice's measurement outcome and decides to perform $\logical{U}$ earlier than intended.  Eventually Alice performs the logical Bell measurement and sends Bob the two classical bits describing the outcome, corresponding to a logical Pauli $\logical{Q}$.  To complete the teleportation procedure, Bob now needs to do something slightly different than before: First, he must undo the $\logical{U}$ he performed prematurely, then perform $\logical{Q}$, and then finally redo $\logical{U}$, now in the correct place.  That is, he should do the gate $\logical{UQU^\dagger}$.  This procedure is pictured in figure~\ref{fig:gateteleport}.
It may not seem like we have gained anything by doing this, but for some special gates $U$, we have.  We can imagine the state $(I \otimes \logical{U}) (\ket{\logical{00}} + \ket{\logical{11}})$ as a special ancilla state, a replacement for the EPR pair normally used in teleportation, and we can prepare it separately.  Since it is a fixed ancilla state, independent of the data, we can apply some special tricks to preparing it.

We still have to perform the gate $\logical{UQU^\dagger}$, which cannot be done ahead of time on the ancilla, since $Q$ depends on the outcome of a logical Bell measurement on the data block.  However, the gate $\logical{UQU^\dagger}$ might be simpler to perform than $\logical{U}$ was.  For instance, when $U \in \Clifford_1$, $UQU^\dagger \in \Pauli_1$ --- that is the defining property of the Clifford group.  For some gates $U \not\in \Clifford_1$, it is nonetheless still true that $UQU^\dagger \in \Clifford_1$ for any $Q \in \Pauli_1$.  For instance, the $\pi/8$ rotation $R_{\pi/8}$ has this property:
\begin{align}
R_{\pi/8} X R_{\pi/8}^\dagger &= \begin{pmatrix} 0 & e^{-i\pi/4} \\ e^{i \pi/4} & 0 \end{pmatrix} = e^{i\pi/4} X P^\dagger \\
R_{\pi/8} Z R_{\pi/8}^\dagger &= Z
\end{align}
We sometimes call the set of unitary operators with this property, of conjugating Pauli operators into Clifford group operators, $C_3$.  $C_1$ is the Pauli group $\Pauli_n$, and $C_2$ is the Clifford group $\Clifford_1$.\footnote{I apologize for the similar appearance of $\Clifford_n$ and $C_k$.  $C_k$ will not appear outside this paragraph.}  One can define a set $C_k = \{ U | UQU^\dagger \in C_{k-1}\ \forall Q \in C_1 \}$, and the teleportation construction tells us how, given appropriate ancilla states, to perform a gate from $C_k$ once we know how to perform gates from $C_{k-1}$.  Note that the sets $C_k$ are not closed under multiplication.

This gives us an indication of how to perform a universal set of fault-tolerant gates.  For the $7$-qubit code, and some similar CSS codes, we already know how to perform all logical Clifford group operations.  The Bell measurement is a Clifford group operation (plus measurement, which we also know how to do fault-tolerantly), and now we have seen that $R_{\pi/8} Q R_{\pi/8}^\dagger$ is also a Clifford group operation for $Q \in \Pauli$.  Thus, if we can just prepare the special ancilla state $(I \otimes \logical{R_{\pi/8}}) (\ket{\logical{00}} + \ket{\logical{11}})$ fault-tolerantly, we can do the complete gate teleportation procedure for the $7$-qubit code to implement a fault-tolerant $\logical{R_{\pi/8}}$ gate.  Since the Clifford group plus $R_{\pi/8}$ form a universal set of gates, that gives us a universal fault-tolerant gate set for the $7$-qubit code.

It is perhaps worth saying as an aside a few words about the nature of this universal set of gates.  Note that for a fixed number of (logical) qubits, it contains a finite set of gates.  These gates do not commute with each other, and if we look at the group generated by these gates, we find that it is an infinite group.  However, the unitary group is uncountable, so we will not be able to exactly implement all unitary gates.  We can, however, get arbitrarily close to any gate we wish, and the approximation procedure can be done quite efficiently as a consequence of the Solovay-Kitaev theorem~\cite{Kitaev:book,Solovay}.  For any gate $U$, to approximate $U$ to within a distance $\epsilon$, we need only use $\text{polylog} (1/\epsilon)$ gates.

Now back to fault tolerance.  I have described the construction above for single-qubit gates $U$, but that is in no way essential.  For a multiple-qubit gate $U$, we simply need, as an ancilla, a state composed of multiple encoded EPR pairs with $U$ applied jointly on the second half of all of them.  For instance, we can perform the $\logical{\text{CNOT}}$ in this way using a four-qubit ancilla.  If we take any stabilizer code with stabilizer $S$, we can perform the Bell measurement as part of Knill error correction, and $N(S)$ gives us transversal implementations of the logical Pauli group.  Thus, using gate teleportation, given an appropriate ancilla, we get a fault-tolerant implementation for any Clifford group gate on $S$. Then we can use gate teleportation again to get a universal set of gates.  Thus, we can perform a universal set of gates for any stabilizer code~\cite{Gottesman:FT}.

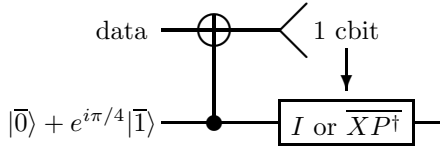
\begin{figure}
\begin{centering}
\begin{picture}(175,55)

\thicklines

\put(30,45){\makebox(30,0){data}}
\put(60,45){\line(1,0){45}}

\put(60,10){\line(1,0){45}}
\put(0,10){\makebox(60,0){$\ket{\logical{0}} + e^{i\pi/4} \ket{\logical{1}}$}}

\put(80,10){\circle*{6}}
\put(80,45){\circle{12}}
\put(80,10){\line(0,1){41}}

\put(105,45){\line(1,1){10}}
\put(105,45){\line(1,-1){10}}

\put(130,45){\makebox(0,0){$1$ cbit}}
\put(130,38){\vector(0,-1){16}}

\put(105,2){\framebox(50,16){$I$ or $\logical{XP^\dagger}$}}
\put(155,10){\line(1,0){10}}

\end{picture}
\caption{More efficient gate teleportation of $\logical{R_{\pi/8}}$.  The CNOT represents a logical $\logical{\text{CNOT}}$, perhaps performed transversally.}
\label{fig:pi8}
\end{centering}
\end{figure}

For certain gates, even simpler constructions are available that take advantage of special properties of the gate~\cite{BMPRV,Shor:FT,ZLC}.  For instance, $R_{\pi/8}$ is a diagonal gate, and that lets us use the construction in figure~\ref{fig:pi8}.
The more efficient construction only involves a single encoded block as ancilla, which is somewhat easier to create.

\subsection{Fault-tolerant state preparation}
\label{sec:prep}

The constructions I have presented in the last few sections rely on the ability to fault-tolerantly prepare some particular states.  It may seem like we have just postponed the main difficulty in performing non-Clifford group gates and fault-tolerant error correction, and to a certain extent this is true.  However, since the ancillas we use in the above procedures are being prepared in \emph{known} states, we have more options as to how to create them than we do for performing an operation (such as the $\pi/8$ rotation) on an unknown data state.

In order to perform Steane or Knill EC, we need to know how to create encoded $\ket{\logical{0}}$ and encoded $\ket{\logical{0}} + \ket{\logical{1}}$ states with few errors in them.  We also, of course, need encoded $\ket{\logical{0}}$ states to initialize the fault-tolerant quantum computation.  Our goal, in particular, is to avoid correlated errors in these states.  The desired states can easily be created using some non-fault-tolerant encoding circuit, but because such circuits inevitably involve gates interacting pairs of qubits within the code block we are constructing, a single error during the encoding circuit can cause correlated errors affecting multiple qubits in the new block.  This would certainly fail to satisfy property Prep A, and could easily fail to satisfy Prep B as well.

One way to create states is to use a version of Shor error correction.  If we add a stabilizer generator corresponding to the $\logical{Z}$ (for $\ket{\logical{0}}$) or $\logical{X}$ (for $\ket{\logical{0}} + \ket{\logical{1}}$), then there is just a single state in the code specified by the new stabilizer, and it is the one we wish to create.  Performing Shor EC to correct an arbitrary state to the one codeword of this new code will then produce a fault-tolerant preparation gadget.

Steane and Knill EC don't give us so immediately methods of state preparation since they need themselves ancillas prepared for the state we are trying to create, but they still point us in the right direction.  Suppose we performed Steane EC on the created state without first verifying the ancilla states.  The result would not be reliable, since the states involved could have multiple correlated errors due to only a single error in the encoding circuit.  However, suppose that if instead of correcting the state based on the error syndrome we measure, we simply use the procedure to \emph{detect} whether errors are present.  Then if we see any errors, we discard the main state we are trying to create as well as the measured ancillas.  In order for a correlated error to slip past this procedure, we need correlated errors both on the data block we are trying to verify and on one or both ancilla blocks, and furthermore, those errors must cancel in the error syndrome.  Since the data block and ancilla blocks are created separately, we need two separate faults for this to be possible: one during the creation of the data block and one during the creation of the ancilla block.  We may need to repeat this procedure using ancillas that themselves have passed previous rounds of screening, but with a sufficient number of iterations, we can fault-tolerantly prepare states even for large distance codes.

For gate teleportation, we need more complicated ancillas.  I will not go into the details, but simply mention some of the main methods used in preparing such states.  For the ancillas needed to teleport Clifford group gates, we can use the method based on Shor EC or similar verification procedures to those for $\ket{\logical{0}}$ or $\ket{\logical{0}} + \ket{\logical{1}}$.  We have the same two options for other gates such as $R_{\pi/8}$, but both methods become more complicated.  The ancilla states used for $R_{\pi/8}$ and similar gates are eigenstates of Clifford group operators, and can be uniquely specified by this fact.  For instance, the state $\ket{\logical{0}} + e^{i\pi/4} \ket{\logical{1}}$ is the unique codeword which is a $+1$ eigenstate of $e^{i\pi/4} \logical{XP^\dagger}$.  Thus, if we measure the eigenvalues of $\logical{XP^\dagger}$, which we can do using a Shor-like method involving a cat state and repetition, we can determine if the ancilla state is correct or not~\cite{Shor:FT}.

There is also a method for verifying this state and some other related ancilla states which involves comparing a number of copies of the state to test if any have errors on them.  As before, we imagine using some non-fault-tolerant procedure to create candidate ancilla states, and then we put them through this verification procedure and discard any states that fail.  The states that come out of the procedure are more likely to be error-free (or at least less likely to have errors dating from the non-fault-tolerant encoding part of the protocol).  We can then take a pool of previously-tested states and iterate the verification procedure to purify them further.  For details, see~\cite{BK}.

\section{Fault-tolerant circuits and the threshold theorem}
\label{sec:threshold}

\subsection{Good and bad extended rectangles}

We have finally shown that there exist gadgets for fault-tolerant state preparation, measurement, error correction, and a universal set of gates.  Now we are ready to study what happens when we apply these gadgets.  I will first show that provided errors do not occur too frequently, we can put together a sequence of fault-tolerant gadgets to create a circuit that gives the same output as the ideal unencoded circuit.

\begin{defn}
Let $C$ be a quantum circuit consisting of a set of locations $C_i$, where each $C_i$ is a preparation location, a measurement location, a gate location, or a wait location.  The preparation locations introduce new qubits into the circuit, and the measurement locations remove qubits from the circuit.  We assume that $C$ can be divided up into time steps so that at each time step, every qubit (not counting those to be added at later time steps or removed at earlier time steps) is involved in exactly one location, that the first location for every qubit is a preparation location, and that the last location for every qubit is a measurement.  Two locations are considered to be \emph{consecutive} if they occur at adjacent time steps and they share a qubit.  (If both are two-qubit gates, they only need to share one qubit.)
The \emph{fault-tolerant protocol for $C$} (using a particular $[[n,1,2t+1]]$ QECC) is a quantum circuit $C'$ constructed by replacing each location $C_i$ with a fault-tolerant gadget for $C_i$, and adding fault-tolerant error correction gadgets between any pair of consecutive locations.  $C'$ is referred to as a \emph{fault-tolerant circuit for $C$} or a \emph{fault-tolerant simulation of $C$}.
\end{defn}

\begin{figure}
\begin{centering}
\begin{picture}(280,70)(0,-5)

\thicklines

% FT preparation
\put(20,45){\oval(20,20)[l]}
\put(20,35){\line(0,1){20}}
\put(20,45){\line(1,0){20}}
\put(23,50){\makebox(0,0)[bl]{$\scriptstyle s_1$}}

% FT preparation
\put(20,15){\oval(20,20)[l]}
\put(20,5){\line(0,1){20}}
\put(20,15){\line(1,0){20}}
\put(23,20){\makebox(0,0)[bl]{$\scriptstyle s_2$}}

% FT EC
\put(40,35){\framebox(20,20){\text{EC}}}
\put(60,45){\line(1,0){20}}
\put(63,50){\makebox(0,0)[bl]{$\scriptstyle s_3$}}

\put(6,2){\dashbox{2}(67,26){}}
\put(6,32){\dashbox{2}(67,26){}}

% FT EC
\put(40,5){\framebox(20,20){\text{EC}}}
\put(60,15){\line(1,0){20}}
\put(63,20){\makebox(0,0)[bl]{$\scriptstyle s_4$}}

% FT gate
\put(95,30){\oval(30,50)}
\put(95,30){\makebox(0,0){CNOT}}
\put(110,45){\line(1,0){20}}
\put(110,15){\line(1,0){20}}
\put(110,50){\makebox(0,0)[bl]{$\scriptstyle s_5$}}

% FT EC
\put(130,35){\framebox(20,20){\text{EC}}}
\put(150,45){\line(1,0){20}}
\put(153,50){\makebox(0,0)[bl]{$\scriptstyle s_6$}}

% FT EC
\put(130,5){\framebox(20,20){\text{EC}}}
\put(150,15){\line(1,0){20}}
\put(153,20){\makebox(0,0)[bl]{$\scriptstyle s_7$}}

\put(36,-2){\dashbox{2}(128,64){}}

% FT gate
\put(180,45){\circle{20}}
\put(180,45){\makebox(0,0){$H$}}
\put(190,45){\line(1,0){20}}
\put(191,50){\makebox(0,0)[bl]{$\scriptstyle s_8$}}

% FT measurement
\put(170,15){\oval(20,20)[r]}
\put(170,5){\line(0,1){20}}
\put(181,20){\makebox(0,0)[bl]{$\scriptstyle s_9$}}

% FT EC
\put(210,35){\framebox(20,20){\text{EC}}}
\put(230,45){\line(1,0){20}}
\put(233,50){\makebox(0,0)[bl]{$\scriptstyle s_{10}$}}

\put(126,2){\dashbox{2}(67,26){}}
\put(125,32){\dashbox{2}(122,26){}}

% FT measurement
\put(250,45){\oval(20,20)[r]}
\put(250,35){\line(0,1){20}}
\put(261,50){\makebox(0,0)[bl]{$\scriptstyle s_{11}$}}

\put(205,28){\dashbox{2}(72,34){}}

\end{picture}
\caption{A schematic representation of a sample fault-tolerant protocol, with extended rectangles marked.}
\label{fig:FTcircuitsample}
\end{centering}
\end{figure}
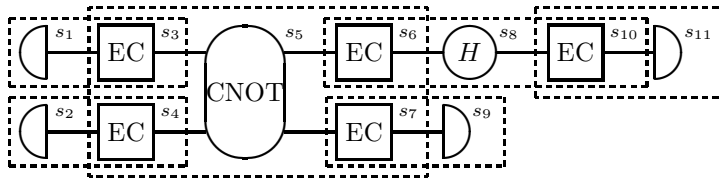

However, it is a bit unwieldy to prove things directly about the circuit as a whole.  Therefore, it makes more sense to focus on smaller units.  We will chop the fault-tolerant circuit up into pieces that contain just a single location from the original circuit $C$:
\begin{defn}
An \emph{extended rectangle} (abbreviated \emph{ExRec}) in the fault-tolerant circuit for $C$ consists of the fault-tolerant gadget corresponding to some location $C_i$ for the original circuit, plus all of the fault-tolerant EC gadgets between $C_i$ and the adjacent locations. We can characterize an ExRec by the type of the location $C_i$ --- e.g., if $C_i$ is a preparation location, then the corresponding extended rectangle is a preparation ExRec.  The EC step(s) before the gadget for $C_i$ are \emph{leading} EC step(s), and those after the gadget for $C_i$ are \emph{trailing} EC step(s).
\end{defn}
Figure~\ref{fig:FTcircuitsample} gives an example of a fault-tolerant circuit, with the ExRecs indicated by dashed lines.  Note that a preparation ExRec or a measurement ExRec only contains a single EC, while a gate (or wait) ExRec contains two EC steps for each qubit involved in the gate.  Except for a preparation ExRec, all ExRecs start with EC steps.  Also, adjacent extended rectangles overlap.  This is a complication, but is actually simpler to deal with than the alternative of having non-overlapping rectangles.  This is because we would like to somehow capture a notion that a rectangle ``behaves correctly'' when it does not have too many faulty locations.  However, if the rectangles do not overlap, there could be an accumulation of errors at the end of one rectangle and at the beginning of the subsequent rectangle.  There is no time to correct any errors between the two rectangles, so those errors would combine to produce more errors than one might na\"{\i}vely expect just counting faulty locations in one rectangle.  This can be dealt with, of course, but that means not only that analyzing the behavior of a single rectangle depends on the behavior of the previous rectangle, but also that we are not taking full advantage of the QECC's ability to correct errors.

So what does it mean to say that an extended rectangle behaves correctly?  It should mean, roughly speaking, that the encoded state after the extended rectangle is the same as encoded state before the extended rectangle but with the appropriate logical operation performed on the state.  However, we still have the complication about having a collection of errors just before the beginning of an ExRec and then more just after the beginning.  That could mean that the state just before the beginning of the ExRec is correctable, but then after the first time step of the ExRec, before error correction has really had a chance to act, it is no longer correctable.  Instead of looking at the state before the beginning of the ExRec, we should instead look at the state after the leading EC step(s) of the ExRec but before the main gadget for the location $C_i$.  By that point, any errors from the previous rectangle have either been corrected or at least combined with errors that occurred during the leading EC steps.  Thus, we make the following definitions:
\begin{defn}
A gate or wait ExRec is \emph{correct} for some particular arrangements of faults if
\begin{equation}
\begin{picture}(155,30)(0,10)

\thicklines

% FT EC
\put(10,15){\line(1,0){10}}
\put(20,5){\framebox(20,20){\text{EC}}}
\put(40,15){\line(1,0){15}}
\put(43,20){\makebox(0,0)[bl]{$\scriptstyle s_1$}}

% FT gate
\put(65,15){\circle{20}}
\put(65,15){\makebox(0,0){$U$}}
\put(75,15){\line(1,0){15}}
\put(76,20){\makebox(0,0)[bl]{$\scriptstyle s_2$}}

% FT EC
\put(90,5){\framebox(20,20){\text{EC}}}
\put(110,15){\line(1,0){15}}
\put(113,20){\makebox(0,0)[bl]{$\scriptstyle s_3$}}

% ideal decoder
\put(125,5){\line(0,1){20}}
\put(125,25){\line(1,-1){10}}
\put(135,15){\line(-1,-1){10}}
\thinlines
\put(135,15){\line(1,0){10}}

\end{picture}
=
\begin{picture}(115,30)(0,10)

\thicklines

% FT EC
\put(10,15){\line(1,0){10}}
\put(20,5){\framebox(20,20){\text{EC}}}
\put(40,15){\line(1,0){15}}
\put(43,20){\makebox(0,0)[bl]{$\scriptstyle s_1$}}

% ideal decoder
\put(55,5){\line(0,1){20}}
\put(55,25){\line(1,-1){10}}
\put(65,15){\line(-1,-1){10}}
\thinlines
\put(65,15){\line(1,0){10}}

% ideal gate
\put(85,15){\circle{20}}
\put(85,15){\makebox(0,0){$U$}}
\put(95,15){\line(1,0){10}}

\end{picture}
\end{equation}
\vspace{5pt}

A preparation ExRec is \emph{correct} for some particular arrangements of faults if
\begin{equation}
\begin{picture}(100,25)
\thicklines

% FT preparation
\put(20,5){\oval(20,20)[l]}
\put(20,-5){\line(0,1){20}}
\put(20,5){\line(1,0){15}}
\put(23,10){\makebox(0,0)[bl]{$\scriptstyle s_1$}}

% FT EC
\put(35,-5){\framebox(20,20){\text{EC}}}
\put(55,5){\line(1,0){15}}
\put(58,10){\makebox(0,0)[bl]{$\scriptstyle s_2$}}

% ideal decoder
\put(70,-5){\line(0,1){20}}
\put(70,15){\line(1,-1){10}}
\put(80,5){\line(-1,-1){10}}
\thinlines
\put(80,5){\line(1,0){10}}

\end{picture}
=
\begin{picture}(40,25)
\thinlines

% ideal preparation
\put(20,5){\oval(20,20)[l]}
\put(20,-5){\line(0,1){20}}
\put(20,5){\line(1,0){10}}

\end{picture}
\end{equation}
\vspace{5pt}

A measurement ExRec is \emph{correct} for some particular arrangement of faults if
\begin{equation}
\begin{picture}(75,30)

\thicklines

% FT EC
\put(10,5){\line(1,0){10}}
\put(20,-5){\framebox(20,20){\text{EC}}}
\put(40,5){\line(1,0){15}}
\put(43,10){\makebox(0,0)[bl]{$\scriptstyle s_1$}}

% FT measurement
\put(55,5){\oval(20,20)[r]}
\put(55,-5){\line(0,1){20}}
\put(66,10){\makebox(0,0)[bl]{$\scriptstyle s_2$}}
\end{picture}
=
\begin{picture}(85,25)
\thicklines

% FT EC
\put(10,5){\line(1,0){10}}
\put(20,-5){\framebox(20,20){\text{EC}}}
\put(40,5){\line(1,0){15}}
\put(43,10){\makebox(0,0)[bl]{$\scriptstyle s_1$}}

% ideal decoder
\put(55,-5){\line(0,1){20}}
\put(55,15){\line(1,-1){10}}
\put(65,5){\line(-1,-1){10}}
\thinlines
\put(65,5){\line(1,0){10}}

% ideal measurement
\put(75,5){\oval(20,20)[r]}
\put(75,-5){\line(0,1){20}}
\end{picture}
\end{equation}
\vspace{5pt}
\end{defn}

We expect that an ExRec should be correct if it contains no more errors than the code can correct.  Thus, we make the following definition:
\begin{defn}
A full ExRec is \emph{good} if it contains at most $t$ faults, where the QECC in use corrects $t$ errors.  A full ExRec is \emph{bad} if it is not good.
\end{defn}
The caveat ``full ExRec'' is needed for a technical reason which I will explain later.  For now, ignore it.

The definition of good is justified by the following theorem:
\begin{thm}[Good implies correct]
\label{thm:correct}
A good ExRec is correct.
\end{thm}

\begin{proof}
The proof is a fairly straightforward application of the definitions of fault-tolerant gadgets.  I will present the case of a gate ExRec.  Measurement and preparation ExRecs work similarly.  Since the ExRec is good, in the following diagrams, $s_1 + s_2 + s_3 \leq t$.

\begin{equation*}
\begin{picture}(280,30)(0,10)

\thicklines

% FT EC
\put(10,15){\line(1,0){10}}
\put(20,5){\framebox(20,20){\text{EC}}}
\put(40,15){\line(1,0){15}}
\put(43,20){\makebox(0,0)[bl]{$\scriptstyle s_1$}}

% FT gate
\put(65,15){\circle{20}}
\put(65,15){\makebox(0,0){$U$}}
\put(75,15){\line(1,0){15}}
\put(76,20){\makebox(0,0)[bl]{$\scriptstyle s_2$}}

% FT EC
\put(90,5){\framebox(20,20){\text{EC}}}
\put(110,15){\line(1,0){15}}
\put(113,20){\makebox(0,0)[bl]{$\scriptstyle s_3$}}

% ideal decoder
\put(125,5){\line(0,1){20}}
\put(125,25){\line(1,-1){10}}
\put(135,15){\line(-1,-1){10}}
\thinlines
\put(135,15){\line(1,0){10}}

\end{picture}
\end{equation*}
\begin{equation}
\begin{picture}(240,30)(-40,10)

\put(-5,15){\makebox(0,0){$=$}}
\thicklines

% FT EC
\put(10,15){\line(1,0){10}}
\put(20,5){\framebox(20,20){\text{EC}}}
\put(40,15){\line(1,0){15}}
\put(43,20){\makebox(0,0)[bl]{$\scriptstyle s_1$}}

% s-filter
\put(55,5){\framebox(5,20){}}
\put(60,15){\line(1,0){15}}
\put(63,20){\makebox(0,0)[bl]{$\scriptstyle s_1$}}

% FT gate
\put(85,15){\circle{20}}
\put(85,15){\makebox(0,0){$U$}}
\put(95,15){\line(1,0){15}}
\put(96,20){\makebox(0,0)[bl]{$\scriptstyle s_2$}}

% FT EC
\put(110,5){\framebox(20,20){\text{EC}}}
\put(130,15){\line(1,0){15}}
\put(133,20){\makebox(0,0)[bl]{$\scriptstyle s_3$}}

% ideal decoder
\put(145,5){\line(0,1){20}}
\put(145,25){\line(1,-1){10}}
\put(155,15){\line(-1,-1){10}}
\thinlines
\put(155,15){\line(1,0){10}}

\put(210,15){\makebox(0,0)[l]{\text{by EC A}}}

\end{picture}
\end{equation}
\begin{equation}
\begin{picture}(240,30)(-40,10)

\put(-5,15){\makebox(0,0){$=$}}
\thicklines

% FT EC
\put(10,15){\line(1,0){10}}
\put(20,5){\framebox(20,20){\text{EC}}}
\put(40,15){\line(1,0){15}}
\put(43,20){\makebox(0,0)[bl]{$\scriptstyle s_1$}}

% s-filter
\put(55,5){\framebox(5,20){}}
\put(60,15){\line(1,0){15}}
\put(63,20){\makebox(0,0)[bl]{$\scriptstyle s_1$}}

% FT gate
\put(85,15){\circle{20}}
\put(85,15){\makebox(0,0){$U$}}
\put(95,15){\line(1,0){15}}
\put(96,20){\makebox(0,0)[bl]{$\scriptstyle s_2$}}

% s-filter
\put(110,5){\framebox(5,20){}}
\put(115,15){\line(1,0){30}}
\put(118,20){\makebox(0,0)[bl]{$\scriptstyle s_1+s_2$}}

% FT EC
\put(145,5){\framebox(20,20){\text{EC}}}
\put(165,15){\line(1,0){15}}
\put(168,20){\makebox(0,0)[bl]{$\scriptstyle s_3$}}

% ideal decoder
\put(180,5){\line(0,1){20}}
\put(180,25){\line(1,-1){10}}
\put(190,15){\line(-1,-1){10}}
\thinlines
\put(190,15){\line(1,0){10}}

\put(210,15){\makebox(0,0)[l]{\text{by Gate A}}}

\end{picture}
\end{equation}
\begin{equation}
\begin{picture}(240,30)(-40,10)

\put(-5,15){\makebox(0,0){$=$}}
\thicklines

% FT EC
\put(10,15){\line(1,0){10}}
\put(20,5){\framebox(20,20){\text{EC}}}
\put(40,15){\line(1,0){15}}
\put(43,20){\makebox(0,0)[bl]{$\scriptstyle s_1$}}

% s-filter
\put(55,5){\framebox(5,20){}}
\put(60,15){\line(1,0){15}}
\put(63,20){\makebox(0,0)[bl]{$\scriptstyle s_1$}}

% FT gate
\put(85,15){\circle{20}}
\put(85,15){\makebox(0,0){$U$}}
\put(95,15){\line(1,0){15}}
\put(96,20){\makebox(0,0)[bl]{$\scriptstyle s_2$}}

% s-filter
\put(110,5){\framebox(5,20){}}
\put(115,15){\line(1,0){30}}
\put(118,20){\makebox(0,0)[bl]{$\scriptstyle s_1+s_2$}}

% ideal decoder
\put(145,5){\line(0,1){20}}
\put(145,25){\line(1,-1){10}}
\put(155,15){\line(-1,-1){10}}
\thinlines
\put(155,15){\line(1,0){10}}

\put(210,15){\makebox(0,0)[l]{\text{by EC B}}}

\end{picture}
\end{equation}
\begin{equation}
\begin{picture}(240,30)(-40,10)

\put(-5,15){\makebox(0,0){$=$}}
\thicklines

% FT EC
\put(10,15){\line(1,0){10}}
\put(20,5){\framebox(20,20){\text{EC}}}
\put(40,15){\line(1,0){15}}
\put(43,20){\makebox(0,0)[bl]{$\scriptstyle s_1$}}

% s-filter
\put(55,5){\framebox(5,20){}}
\put(60,15){\line(1,0){15}}
\put(63,20){\makebox(0,0)[bl]{$\scriptstyle s_1$}}

% FT gate
\put(85,15){\circle{20}}
\put(85,15){\makebox(0,0){$U$}}
\put(95,15){\line(1,0){15}}
\put(96,20){\makebox(0,0)[bl]{$\scriptstyle s_2$}}

% ideal decoder
\put(110,5){\line(0,1){20}}
\put(110,25){\line(1,-1){10}}
\put(120,15){\line(-1,-1){10}}
\thinlines
\put(120,15){\line(1,0){10}}

\put(210,15){\makebox(0,0)[l]{\text{by Gate A}}}

\end{picture}
\end{equation}
\begin{equation}
\begin{picture}(240,30)(-40,10)

\put(-5,15){\makebox(0,0){$=$}}
\thicklines

% FT EC
\put(10,15){\line(1,0){10}}
\put(20,5){\framebox(20,20){\text{EC}}}
\put(40,15){\line(1,0){15}}
\put(43,20){\makebox(0,0)[bl]{$\scriptstyle s_1$}}

% s-filter
\put(55,5){\framebox(5,20){}}
\put(60,15){\line(1,0){15}}
\put(63,20){\makebox(0,0)[bl]{$\scriptstyle s_1$}}

% ideal decoder
\put(75,5){\line(0,1){20}}
\put(75,25){\line(1,-1){10}}
\put(85,15){\line(-1,-1){10}}
\thinlines
\put(85,15){\line(1,0){15}}

% FT gate
\put(110,15){\circle{20}}
\put(110,15){\makebox(0,0){$U$}}
\put(120,15){\line(1,0){10}}

\put(210,15){\makebox(0,0)[l]{\text{by Gate B}}}

\end{picture}
\end{equation}
\begin{equation}
\begin{picture}(240,30)(-40,10)

\put(-5,15){\makebox(0,0){$=$}}
\thicklines

% FT EC
\put(10,15){\line(1,0){10}}
\put(20,5){\framebox(20,20){\text{EC}}}
\put(40,15){\line(1,0){15}}
\put(43,20){\makebox(0,0)[bl]{$\scriptstyle s_1$}}

% ideal decoder
\put(55,5){\line(0,1){20}}
\put(55,25){\line(1,-1){10}}
\put(65,15){\line(-1,-1){10}}
\thinlines
\put(65,15){\line(1,0){15}}

% FT gate
\put(90,15){\circle{20}}
\put(90,15){\makebox(0,0){$U$}}
\put(100,15){\line(1,0){10}}

\put(210,15){\makebox(0,0)[l]{\text{by EC A}}}

\end{picture}
\end{equation}

\end{proof}

\subsection{Level reduction}

Now let us consider the case of a complete fault-tolerant circuit.  Suppose \emph{all} the extended rectangles in the circuit are good.  Then by applying theorem~\ref{thm:correct} repeatedly, we can create ideal decoders using the correctness property for the measurement locations, push the ideal decoders all the way to the left using correctness for the gate and wait locations, and then eliminate the ideal decoders using correctness for the preparation locations.  We are left, as desired, with the original circuit $C$ which the fault-tolerant circuit is supposed to simulate.
\begin{thm}
\label{thm:goodcircuit}
Suppose a fault-tolerant circuit for $C$ contains only good extended rectangles.  Then the output distribution of the fault-tolerant protocol is the same as the output distribution of $C$.
\end{thm}

That's precisely the result we wanted, but unfortunately it only applies when all the extended rectangles are good.  In a little bit, once we talk more concretely about the model of errors, you will see that this result is enough to say that a fault-tolerant protocol helps protect against errors, in that, for sufficiently low error rate, the probability of having only good ExRecs in the complete fault-tolerant circuit is greater than the probability of getting through the original circuit $C$ without any errors.  Unfortunately, when $C$ is large, it is still unlikely we can make it all the way through even a fault-tolerant circuit without having a bad extended rectangle unless we go to codes which can correct many errors.  But if we do use codes with large $t$, we will need a special family of codes, since otherwise we have no good way to make the large ancillas we need for fault-tolerant gate constructions and fault-tolerant error correction.  Luckily, there are two known families of codes for which we can solve this problem, giving a threshold theorem.  One of them is based on topological constructions~\cite{DKLP}.  I will consider the other, which is somewhat simpler to analyze: the family of concatenated codes.

The logic behind a concatenated code is quite straightforward.  We have shown (up to a question of defining the right error model) that we can take any circuit $C$ and create a fault-tolerant circuit $C'$ that gives the same result and is more reliable than $C$.  Why not, therefore, take $C'$ and create a fault-tolerant circuit $C''$ simulating it?  Surely that should be even \emph{more} reliable.  And if that is not enough, we can do it again and again until the circuit is robust enough to give us the right answer, even if the original circuit $C$ is very large.  And indeed, this strategy works.  We can create a sequence of circuits $C^{(k)}$, $k = 1, \dots, L$, where each $C^{(k)}$ is a fault-tolerant simulation of $C^{(k-1)}$, with $C^{(0)} = C$.  When $C$ is large, it is not very likely that $C^{(L)}$ contains only good extended rectangles, but it will be true that it simulates a $C^{(L-1)}$ which contains (with high probability, when the error rate is low enough, \dots) fewer bad extended rectangles than $C^{(L)}$, and so on, until, with very high probability, $C^{(1)}$ does contain only good extended rectangles, and thus gives us a nearly perfect simulation of $C$.

In order to prove this, however, we will need to understand what happens to a bad extended rectangle when we push an ideal decoder through it from right to left.  We would like to say that a bad extended rectangle simulates a faulty location in the original circuit, and this is almost true.  In particular, we would like for the following to be true when $s_1 + s_2 + s_3 > t$, for some erroneous gate $U'$:
\begin{equation}
\begin{picture}(155,30)(0,10)

\thicklines

% FT EC
\put(10,15){\line(1,0){10}}
\put(20,5){\framebox(20,20){\text{EC}}}
\put(40,15){\line(1,0){15}}
\put(43,20){\makebox(0,0)[bl]{$\scriptstyle s_1$}}

% FT gate
\put(65,15){\circle{20}}
\put(65,15){\makebox(0,0){$U$}}
\put(75,15){\line(1,0){15}}
\put(76,20){\makebox(0,0)[bl]{$\scriptstyle s_2$}}

% FT EC
\put(90,5){\framebox(20,20){\text{EC}}}
\put(110,15){\line(1,0){15}}
\put(113,20){\makebox(0,0)[bl]{$\scriptstyle s_3$}}

% ideal decoder
\put(125,5){\line(0,1){20}}
\put(125,25){\line(1,-1){10}}
\put(135,15){\line(-1,-1){10}}
\thinlines
\put(135,15){\line(1,0){10}}

\end{picture}
=
\begin{picture}(115,30)(0,10)

\thicklines

% FT EC
\put(10,15){\line(1,0){10}}
\put(20,5){\framebox(20,20){\text{EC}}}
\put(40,15){\line(1,0){15}}
\put(43,20){\makebox(0,0)[bl]{$\scriptstyle s_1$}}

% ideal decoder
\put(55,5){\line(0,1){20}}
\put(55,25){\line(1,-1){10}}
\put(65,15){\line(-1,-1){10}}
\thinlines
\put(65,15){\line(1,0){10}}

% faulty gate
\put(85,15){\circle{20}}
\put(85,15){\makebox(0,0){$U'$}}
\put(95,15){\line(1,0){10}}

\end{picture}
\label{eq:wrongbad}
\vspace{10pt}
\end{equation}
Unfortunately, we cannot easily make any such statement.  The problem is that the error that occurs on the physical gate after the ideal decoder can depend not only on the faults in the gate gadget and trailing EC, but also on the faults in the leading EC and even the carryover error from the previous ExRec, so there is no way to define $U'$ without looking at the larger context.  To handle this problem, we introduce an alternate ideal decoder, a \emph{*-decoder}, which keeps track of the error syndrome:
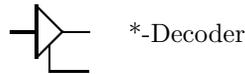
\begin{equation*}
\begin{picture}(120,40)(90,0)

\thicklines

% *-decoder
\put(100,20){\line(1,0){10}}
\put(110,10){\line(0,1){20}}
\put(110,30){\line(1,-1){10}}
\put(120,20){\line(-1,-1){10}}
\thinlines
\put(120,20){\line(1,0){10}}
\put(115,15){\line(0,-1){10}}
\put(115,5){\line(1,0){15}}

\put(145,12){\makebox(55,16)[l]{\text{*-Decoder}}}

\end{picture}
\end{equation*}
The line dropping out the bottom of the *-decoder contains the information about the error syndrome of the state input on the left.  Consequently, the *-decoder is a unitary operation.  To replace (\ref{eq:wrongbad}), we thus can write:
\begin{equation}
\begin{picture}(155,30)(0,10)

\thicklines

% FT EC
\put(10,15){\line(1,0){10}}
\put(20,5){\framebox(20,20){\text{EC}}}
\put(40,15){\line(1,0){15}}
\put(43,20){\makebox(0,0)[bl]{$\scriptstyle s_1$}}

% FT gate
\put(65,15){\circle{20}}
\put(65,15){\makebox(0,0){$U$}}
\put(75,15){\line(1,0){15}}
\put(76,20){\makebox(0,0)[bl]{$\scriptstyle s_2$}}

% FT EC
\put(90,5){\framebox(20,20){\text{EC}}}
\put(110,15){\line(1,0){15}}
\put(113,20){\makebox(0,0)[bl]{$\scriptstyle s_3$}}

% ideal *-decoder
\put(125,5){\line(0,1){20}}
\put(125,25){\line(1,-1){10}}
\put(135,15){\line(-1,-1){10}}
\thinlines
\put(135,15){\line(1,0){10}}
\put(130,10){\line(0,-1){10}}
\put(130,0){\line(1,0){15}}

\end{picture}
=
\begin{picture}(115,30)(35,10)

\thicklines

% ideal *-decoder
\put(45,15){\line(1,0){10}}
\put(55,5){\line(0,1){20}}
\put(55,25){\line(1,-1){10}}
\put(65,15){\line(-1,-1){10}}
\thinlines
\put(65,15){\line(1,0){10}}
\put(60,10){\line(0,-1){10}}
\put(60,0){\line(1,0){15}}

% faulty gate
\put(85,10){\oval(20,30)}
\put(85,10){\makebox(0,0){$U'$}}
\put(95,15){\line(1,0){10}}
\put(95,0){\line(1,0){10}}

\end{picture}
\label{eq:rightbad}
\vspace{20pt}
\end{equation}
The erroneous gate $U'$ uses the error syndrome as a control to tell it what sort of error to apply to the decoded qubit.
Notice that we have pushed the *-decoder all the way through the ExRec, eliminating the leading EC as well as the gate gadget and trailing EC.  This is needed so that bad ExRecs do not overlap, since that could produce correlated errors in the circuit being simulated.  I shall return to this point shortly.

Correctness is defined using *-decoders in a way that is essentially identical to using regular ideal decoders.  The only difference is that we must account for what happens to the error syndrome.  Even if we move a *-decoder through a gadget with no faults, the error syndrome produced will generally change to account for any gates performed on the erroneous physical qubits.  For instance, if we perform a transversal Hadamard, pre-existing physical $X$ errors will change to physical $Z$ errors, which have a different error syndrome.  For this reason, we can define correctness for *-decoders using identical diagrams to those for regular ideal decoders, but with the error syndrome changed in some way, which we do not need to specify.  I.e., a gate ExRec is correct for *-decoders if it satisfies
\begin{equation}
\begin{picture}(155,30)(0,10)

\thicklines

% FT EC
\put(10,15){\line(1,0){10}}
\put(20,5){\framebox(20,20){\text{EC}}}
\put(40,15){\line(1,0){15}}
\put(43,20){\makebox(0,0)[bl]{$\scriptstyle s_1$}}

% FT gate
\put(65,15){\circle{20}}
\put(65,15){\makebox(0,0){$U$}}
\put(75,15){\line(1,0){15}}
\put(76,20){\makebox(0,0)[bl]{$\scriptstyle s_2$}}

% FT EC
\put(90,5){\framebox(20,20){\text{EC}}}
\put(110,15){\line(1,0){15}}
\put(113,20){\makebox(0,0)[bl]{$\scriptstyle s_3$}}

% ideal *-decoder
\put(125,5){\line(0,1){20}}
\put(125,25){\line(1,-1){10}}
\put(135,15){\line(-1,-1){10}}
\thinlines
\put(135,15){\line(1,0){10}}
\put(130,10){\line(0,-1){10}}
\put(130,0){\line(1,0){15}}

\end{picture}
=
\begin{picture}(115,30)(0,10)

\thicklines

% FT EC
\put(10,15){\line(1,0){10}}
\put(20,5){\framebox(20,20){\text{EC}}}
\put(40,15){\line(1,0){15}}
\put(43,20){\makebox(0,0)[bl]{$\scriptstyle s_1$}}

% ideal *-decoder
\put(55,5){\line(0,1){20}}
\put(55,25){\line(1,-1){10}}
\put(65,15){\line(-1,-1){10}}
\thinlines
\put(65,15){\line(1,0){10}}
\put(60,10){\line(0,-1){15}}
\put(60,-5){\line(1,0){15}}

% ideal gate
\put(85,15){\circle{20}}
\put(85,15){\makebox(0,0){$U$}}
\put(95,15){\line(1,0){10}}

% syndrome modification
\put(75,-13){\framebox(20,12){$V$}}
\put(95,-5){\line(1,0){10}}

\end{picture}
\vspace{25pt}
\end{equation}
for some operation $V$ on the error syndrome.  ($V$ of course depends on the errors in the original ExRec as well as the gate $U$.)  The definitions for preparation and measurement ExRecs are similar; for the preparation ExRec, the error syndrome is brought in as a separate input on the RHS of the definition, and for the measurement ExRec, the error syndrome is discarded immediately after being produced.  We then get the following lemma:
\begin{lemma}
\label{lemma:stardecoder}
If an ExRec is correct for an ideal decoder, it is correct for a *-decoder.
\end{lemma}

\begin{proof}
Note that the ideal decoder is just the *-decoder with the error syndrome discarded.  The definition of correctness for an ideal decoder must apply for all input states, including superpositions and parts of entangled states.  This could not hold if the error syndrome output of the *-decoder had any further interaction with the data block after the action of the *-decoder.
\end{proof}

Thus, we also know that good extended rectangles are correct for *-decoders.  We are almost ready to revise theorem~\ref{thm:goodcircuit} to take into account bad ExRecs as well as good ones, but there is one more complication.  We would like to replace decoders with *-decoders in the proof of theorem~\ref{thm:goodcircuit} and push the *-decoders all the way through the circuit.  However, I have declared that a *-decoder should get pushed all the way through a bad ExRec to eliminate the leading EC step.  That means that if we push a *-decoder backwards through a bad ExRec, the previous ExRec will be left without a trailing EC, and will no longer be an ExRec.

\begin{defn}
An ExRec missing one or more trailing ECs (in the case of a multiple-qubit gate ExRec) is called a \emph{truncated} ExRec.  A truncated ExRec is \emph{good} if it contains at most $t$ faulty locations and is \emph{bad} if it is not good.  Within a larger circuit, determine whether an ExRec should be good or bad and full or truncated by the following procedure: Start from the end of the circuit, with measurement ExRecs, and determine whether they are good or bad.  For each other ExRec in the circuit, truncate it by eliminating a trailing EC step if that EC step participates as a leading EC in a bad ExRec (full or truncated).  Once we know whether the ExRec is truncated (on all its output blocks, if there is more than one), determine whether it is good or bad.  We can then determine recursively the nature of all ExRecs in the circuit.
\end{defn}

It is straightforward to define correctness for both regular ideal decoders and *-decoders for truncated ExRecs by just removing the trailing EC step from each diagram.
\begin{lemma}
A good truncated ExRec is correct for both ideal decoders and *-decoders.
\end{lemma}
\begin{proof}
Since the ideal decoder is supposed to incorporate a perfect error correction, we could insert an EC gadget with $0$ errors before any ideal decoder without changing the ouput at all.  (We cannot do this before a *-decoder since the perfect EC gadget would clear the error syndrome.)  Therefore, the correctness diagram for ideal decoders is effectively the same for full and truncated ExRecs, and it follows that a good truncated ExRec is correct for ideal decoders.  Then, applying lemma~\ref{lemma:stardecoder} (which also works for truncated ExRecs), we find that a good truncated ExRec is correct for *-decoders as well.
\end{proof}

That leads us to the following improvement of theorem~\ref{thm:goodcircuit}:
\begin{thm}
\label{thm:goodbadcircuit}
Suppose we have a fault-tolerant circuit $C'$ for $C$.  Assign good and bad extended rectangles to $C'$, and produce a circuit $\tilde{C}$ as follows: If the ExRec for $C_i$ is good, include $C_i$ unchanged in $\tilde{C}$.  If the ExRec for $C_i$ is bad, replace $C_i$ by the erroneous gate $U'$ from eq.~(\ref{eq:rightbad}) or the corresponding equation for the correct type of location.  The circuit $\tilde{C}$ uses ancilla registers to control the types of $U'$ errors.  Then the output distribution of $C'$ is the same as the output distribution of $\tilde{C}$.
\end{thm}
That is, $C'$ simulates a version of $C$ with errors in place of the bad extended rectangles.  We prove this theorem in the same way as theorem~\ref{thm:goodcircuit}, by creating *-decoders at the right end of the circuit and pushing them back to the left using the correctness conditions and eq.~(\ref{eq:rightbad}).  The ancilla registers used to determine the type of errors are the error syndrome registers produced by the *-decoders.

Finally, we are ready to talk about a concrete error model and to prove that fault tolerance reduces the error rate.
\begin{defn}
An \emph{uncorrelated error model} assumes that each location $C_i$ is faulty with probability $p_i$, independently for different locations, and if there is an error at location $C_i$, it is chosen from some distribution $E_i$ independent of the other locations.  In an \emph{uncorrelated Pauli error model}, if a location has a fault, there is probability $p_{Qi}$ of Pauli error $Q$ acting on the qubit(s) involved in the location (relative to the correct operation for that location).  Usually, we assume that $p_i$ and $p_{Qi}$ only depend on what type of location $C_i$ is, and nothing else about it.  Often, we assume that $p_i = p$ for all $i$.  An \emph{uncorrelated depolarizing error model} is a special case of the uncorrelated Pauli error model where $p_{Qi}$ does not depend on $Q$ (except for $Q=I$, no error, which can be different).
\end{defn}
Uncorrelated Pauli error models, and particularly uncorrelated depolarizing error models, are very convenient if one is doing simulations of fault-tolerant protocols to analyze their behavior and error tolerance.  One reason for this is that EC steps for CSS and general stabilizer codes can be performed using only Clifford group gates, and thus the behavior of Pauli errors on a fault-tolerant circuit involving only logical Clifford group gates can be efficiently simulated on a classical computer.  Under more general errors, the simulations quickly become unwieldy after only a few steps.

However, any kind of uncorrelated error model is not going to be quite enough for us.  The reason is that in theorem~\ref{thm:goodbadcircuit}, even if the errors in $C$ are uncorrelated, the location of good and bad ExRecs is somewhat correlated (because the ExRecs overlap, and whether an ExRec is truncated or not depends on whether the following ExRec(s) are good or bad).  Furthermore, the \emph{type} of errors in $\tilde{C}$ are even more correlated, even entangled, because the type of error on a location $\tilde{C}_i$ of $\tilde{C}$ depends on the persistent error syndrome register produced by the *-decoder.  Therefore, even if $C'$ has a simple uncorrelated error model, Pauli or otherwise, $\tilde{C}$ will not.  Instead, we have to go to a slightly more complicated type of error model.

\begin{defn}
Consider a full circuit $C$ and suppose with probability $p_S$, there are faults at precisely the set $S$ of locations, in which case the error at those locations can be any quantum operation consistent with the causal structure of $S$ (i.e., if location $C_i$ is chronologically before $C_j$, the error must act on $C_i$ before $C_j$).  Let $p_i < 1$ be fixed for each location $C_i$.  For any set $R$ of locations of $C$, suppose that the total probability of having faults on every location in $R$ (and possibly additional locations outside $R$) is at most $\prod_{i \in R} p_i$.  Then we have a \emph{local stochastic error model}.
\end{defn}
As before, we usually consider the case where $p_i$ only depends on the type of location $C_i$, and often specialize to the case where $p_i = p$ for all $i$.  In that case, the probability of having errors on all of a specified set of $r$ locations is at most $p^r$.  Note that for this condition, we do not care what happens to locations outside the set $R$ --- some may have errors, others may not.  In addition, we make no constraint on the \emph{type} of error that occurs at any given location $C_i$ with $i \in R$.  Often, we imagine that the type of error is chosen by an adversary who is determined to make our lives as difficult as possible.  The adversary can choose errors that entangle different faulty locations; she may even choose to turn off the errors at some subset of locations if that will cause more trouble than leaving them on.  However, the adversary is restricted to choose a strategy that has the probability of error on any given set of locations decreasing exponentially with the size of the set; it is in this sense that the error model is ``local.''  There is another important restriction which is implicit in the above definition --- we have assumed that the locations of the faults are chosen randomly, which is why it is a ``stochastic'' error model.  This is not the most general possibility.  We could have had a superposition of faults in different sets of locations, perhaps entangled with some environment qubits with persistent memory.  It is possible to generalize the threshold theorem (theorem~\ref{thm:threshold}) to that case, but the details are somewhat complicated~\cite{AGP}.

An uncorrelated error model will automatically be a local stochastic error model, with the same $p_i$.  However, as noted above, if $C'$ is subject to an uncorrelated error model, then it does not generally follow that $\tilde{C}$ will experience an uncorrelated error model as well.  That is the advantage of generalizing to a local stochastic model: If $C'$ is subject to a local stochastic error model, then $\tilde{C}$ will also experience a local stochastic error model.
\begin{thm}[Level Reduction]
\label{thm:levelreduction}
Suppose we have a fault-tolerant circuit $C'$ for $C$, and suppose $C'$ experiences a local stochastic error model with error bounds $p_i$ and probability $p_S$ of errors at precisely the set $S$ of locations.  Then for any particular set $S$ define $\tilde{C}_S$ as in theorem~\ref{thm:goodbadcircuit}, and define an error model on $C$ by replacing $C$ with $\tilde{C}_S$ with probability $p_S$.  This is a local stochastic error model.  The error bounds $p'_i$ for $C$ are given by
\begin{equation}
p'_i \leq \sum_R \prod_{i \in R} p_i.
\label{eq:levelreductionfull}
\end{equation}
The sum over $R$ is taken over sets $R$ of locations which are included in the ExRec for location $C_i$ and with $|R| = t+1$.  (Recall that $t$ is the number of errors the QECC corrects.)  In the special case where $p_i \leq p$ for all $i$, we have
\begin{equation}
p'_i \leq A p^{t+1},
\label{eq:levelreduction}
\end{equation}
where $A$ is the maximum over types of ExRecs of the number of sets of exactly $t+1$ locations in the ExRec.
\end{thm}

\begin{proof}
By theorem~\ref{thm:goodbadcircuit}, we know that the error model for $C$ has a stochastic form, namely that for any set of locations $S'$, there is some probability $p'_{S'}$ of having faults at exactly the locations of $S'$.  To calculate a bound on the probability of having errors at a set $R$ of locations of $C$ (and possibly elsewhere), we should add up the probabilities $p_S$ of every set $S$ of locations of $C'$ which leads to bad ExRecs for locations $C_i$, $i \in R$.

For a single ExRec (say for the location $C_i$), we can upper bound the probability that it is bad by summing over all sets $S$ which contain a subset $R$ of locations which are included in the ExRec and have $|R| = t+1$.  (Any set of locations with $|R| > t+1$ includes as a subset a number of sets with $|R| = t+1$, and is therefore already included in this sum.)  By the union bound and the definition of a local stochastic error model, we therefore know that the probability of a single ExRec being bad is at most $p'_i$, with $p'_i$ given by (\ref{eq:levelreductionfull}).

For a set of ExRecs given by the locations $C_i$, $i \in R'$, we sum over sets $S$ which contain subsets of $t+1$ locations in every ExRec in the set. If the ExRecs do not overlap, we get exactly the bound $\prod_{i \in R'} p'_i$ in this way, considering $S$ to contain a collection of sets $R_i$, $|R_i| = t+1$, $R_i$ a subset of the ExRec for $C_i$.  When the ExRecs overlap, we have to be slightly more careful, since the earlier ExRec is truncated for those cases where the later ExRec is bad, but that just means we sum over $R_i$ which are contained in the truncated ExRec.  To get back to $\prod_{i \in R'} p'_i$, we must add in extra sets of locations which include $R_i$ intersecting the truncated EC step; this can only increase the bound, which is acceptable.
\end{proof}

It is for this theorem that we needed to truncate extended rectangles.  When two ExRecs are completely separate, then in order for \emph{both} to be bad, we need $t+1$ errors in each, and therefore the probability of having both be bad is at most $p^{2(t+1)} = (p'_i)^2$ when $p_i \leq p$.  However, for two overlapping rectangles, if we do not truncate, only $t+1$ errors are needed for both to be bad if all the errors are in the shared EC step.  Then the probability of both failing would be $O(p'_i)$, not $O((p'_i)^2)$.

With some additional computational effort, we can actually set a tighter bound on $p'_i$.  While we have defined an ExRec to be bad when it has $t+1$ or more faulty locations, there are some sets of $t+1$ locations (or more) for which the ExRec remains correct.  One can define a \emph{malignant} set of locations $R$ to be a set for which the ExRec is not correct for some set of errors at $R$.  Then the sum in equation~(\ref{eq:levelreductionfull}) can be taken just over the minimal malignant sets of errors.

Note that, when $p_i = p$, the error rate $p'_i$ for $C$ is less than the physical error rate $p$ for $C'$ if $A p^t < 1$.  In that case, we can apply the idea of concatenated coding to make the logical error rate arbitrarily small.  The threshold theorem then follows easily from theorem~\ref{thm:levelreduction}:

\begin{thm}
\label{thm:threshold}
There is a threshold error rate $p_T$.  Suppose we have a local stochastic error model with $p_i \leq p < p_T$.  Then for any ideal circuit $C$, and any $\epsilon > 0$, there exists a fault-tolerant circuit $C'$ which, when it undergoes the error model, produces an output which has statistical distance at most $\epsilon$ from the output of $C$.  $C'$ has a number of qubits and a number of timesteps which are at most $\text{polylog} (|C|/\epsilon)$ times bigger than the number of qubits and timesteps in $C$, where $|C|$ is the number of locations in $C$.
\end{thm}

\begin{proof}
As noted above, we use concatenated codes.  We take an $[[n,1,2t+1]]$ QECC and create a sequence of circuits $C^{(k)}$, $k = 1, \dots, L$, where each $C^{(k)}$ is a fault-tolerant simulation of $C^{(k-1)}$, with $C^{(0)} = C$.  $C' = C^{(L)}$ undergoes a local stochastic error model, and by theorem~\ref{thm:levelreduction}, so does $C^{(k)}$ for all $k < L$.  The local stochastic error model for $C^{(k)}$ has error bound $p_i^{(L-k)} \leq p^{(L-k)}$, with $p^{(0)} = p$ and
\begin{equation}
p^{(j)} \leq A \left(p^{(j-1)}\right)^{t+1} = p^{(j-1)} \left( p^{(j-1)}/p_T \right)^{t},
\end{equation}
with $p_T = 1/A^{1/t}$.  It follows that $p^{(1)}/p_T \leq (p/p_T)^{t+1}$, $p^{(2)}/p_T \leq (p/p_T)^{(t+1)^2}$, and
\begin{equation}
p^{(j)}/p_T \leq (p/p_T)^{(t+1)^j}.
\end{equation}
The logical error rate after $L$ levels of encoding thus decreases with $L$ as a double exponential when $p < p_T$.

If we wish to achieve a final error rate of $\epsilon_1$ per location in $C$, we therefore need to choose
\begin{equation}
L = \left\lceil \log_{t+1} \left[ \log (\epsilon_1/p_T)/ \log (p/p_T) \right] \right\rceil.
\label{eq:levelsneeded}
\end{equation}
That is, for this choice of $L$, we find that $C$ undergoes a local stochastic error model with error bound $p_i \leq \epsilon_1$; this gives an upper bound on the probability of having a fault on location $i$.  It therefore follows that the probability of having an error in some location of $C$ is at most $\epsilon = |C| \epsilon_1$.  With probability $1-\epsilon$ there are no faults in the simulated circuit $C$, and therefore the overall statistical difference of the output is at most $\epsilon$ from the correct output distribution.

The size of $C'$ is given by $L$.  The total number of qubits, including ancillas, involved in a fault-tolerant gadget is at most some constant $G$, as is the total number of time steps.  Thus, the number of qubits and time steps involved in $C^{(k)}$ is at most $G$ times the number of qubits and time steps involved in $C^{(k-1)}$, and in particular, the number of qubits/time steps in $C^{(L)}$ is at most $G^L$ times the number of qubits/time steps in $C$.  Choosing $L$ as in equation~(\ref{eq:levelsneeded}), we find
\begin{equation}
G^L \leq G \left[ \log (\epsilon_1/p_T)/ \log (p/p_T) \right]^{\log G/\log (t+1)}.
\end{equation}
\end{proof}

To be more explicit, we can choose a particular $[[n,1,2t+1]]$ QECC and fault-tolerant protocol, and for that protocol determine $A$ in equation~(\ref{eq:levelreduction}).  Then the threshold for that protocol is at least $p_T = 1/A^{1/t}$.  By choosing a different code or protocol, or a different method of analysis, we might get a higher threshold, and the true threshold $p_T$ is the supremum over all possible choices.  If we use the $7$-qubit code and are careful counting malignant sets of errors, we find $p_T \geq 2.73 \times 10^{-5}$~\cite{AGP,Reichardt:d3}.  People have studied a variety of different codes and fault-tolerant protocols~\cite{CDT}.  So far the best claimed thresholds have come in simulations by Knill~\cite{Knill:EC}, who has used elaborate ancilla preparation techniques to achieve a threshold $p_T$ of as high as $5\%$, depending on the details of the error model.  The current best rigorous lower bound on the threshold, using those same techniques, gives a threshold of $10^{-3}$~\cite{AGP:d2,Reichardt:thesis}.

The threshold theorem can be improved in various ways.  A similar theorem can be proven using more general noise produced by a weak interaction between the computer and a non-Markovian environment~\cite{AGP,TB}.  This includes systematic errors --- for instance, if every time we perform a $R_\theta$ phase rotation, we over-rotate by a consistent small angle.  The gadgets presented in this paper assume we can perform gates between arbitrary pairs of qubits, but it is also possible to devise gadgets which only involve nearest-neighbor interactions in one or two dimensions~\cite{AB:Threshold,Gottesman:local,RH,SFH,SDT}.  Most quantum gates have a tendency to mix different types of Pauli errors, but if you are careful, it is possible to design fault-tolerant protocols which can take advantage of a large asymmetry between $X$ and $Z$ errors~\cite{AP:phase}.  When you attempt to optimize the threshold, it is generally at a large constant factor cost in overhead.  There has been some study of the tradeoff between overhead and threshold value~\cite{AP:Fibonacci,CDT,Steane:overhead}, but much more could be done in that direction.  Recent work indicates that thresholds can be improved by not treating each level of concatenation separately, but by allowing the EC gadgets to use error information generated by the lower-level error correction procedures~\cite{AP:Fibonacci,ES,Poulin:messagepassing}.

Naturally, there are some assumptions inherent in theorem~\ref{thm:threshold} that cannot be removed.  As long as the physical wait location has non-zero errors, we cannot let any qubit wait around a long time without being corrected.  That means that we must be able to perform operations in parallel; otherwise as the computer gets large, we could only correct errors on a vanishingly small fraction of the qubits at any given time.  It also means we must be able to prepare qubits during the computation~\cite{ABIN}.  This is because we need to use ancilla qubits for error correction, and if the ancilla qubits were prepared only at the beginning of the computation, they would have accumulated a large error rate by the time they are used in an EC gadget.  We must also assume the errors do not remove qubits from our computer irrevocably; either an error takes us to a valid physical computational state (albeit an incorrect one) or at least takes us to a state which can be restored by an appropriate action to some valid computational state.  Finally, we need some sort of bound on the correlations present in the error model.  The local stochastic model allows very strong correlations in the errors, but at least assumes that the probability of having an error on a particular large set of qubits decreases exponentially.  This is roughly what we need in general; the non-Markovian model replaces this assumption with the assumption, roughly speaking, that the \emph{amplitude} of a many-qubit error decreases exponentially.  If instead we had merely a polynomial decay in the amplitude of many-qubit errors, there would be a polynomially-small chance that every qubit in the computer would fail simultaneously, and there is no way we could recover from such a big failure.

\end{document}